\documentclass[conference, 11pt,onecolumn,letterpaper]{IEEEtran}
\makeatletter
\let\IEEEproof\proof
\let\IEEEendproof\endproof
\let\proof\@undefined
\let\endproof\@undefined
\makeatother

\usepackage{amsthm,amssymb,euscript,amscd,amsbsy,psfrag,cite}
\usepackage{graphicx}
\usepackage{graphics}
\usepackage{epsf,subfigure}

\newtheorem{lemma}{Lemma}
\newtheorem{theorem}{Theorem}
\newtheorem{prop}{Proposition}
\newtheorem{definition}{Definition}
\newtheorem{remark}{Remark \textrm}

\newcommand{\disp}{\displaystyle}

\newcommand{\E}{\mathbf{E}}
\newcommand{\trace}{\textnormal{trace}}

\newcommand{\tr}{\textnormal{tr}}

\newcommand{\MMSE}{\mbox{MMSE}}
\newcommand{\MSE}{\mbox{MSE}}
\newcommand{\CMMSE}{\mbox{CMMSE}}
\newcommand{\CRB}{\mbox{CRB}}

\newcommand{\eqa}{\stackrel{\textnormal{(a)}}{=}}
\newcommand{\eqb}{\stackrel{\textnormal{(b)}}{=}}

\newcommand{\leqb}{\stackrel{\textnormal{(b)}}{\leq}}
\newcommand{\leqc}{\stackrel{\textnormal{(c)}}{\leq}}

\newcommand{\bfa}{\pmb{a}}

\newcommand{\bbC}{\mathbb{C}}
\newcommand{\bbA}{\mathbb{A}}
\newcommand{\bbX}{\mathbb{X}}
\newcommand{\bbR}{\mathbb{R}}
\newcommand{\bbD}{\mathbb{D}}

\newcommand{\bbS}{\mathbb{S}}
\newcommand{\bbT}{\mathbb{T}}

\newcommand{\ubbC}{\underline{\mathbb{C}}}
\newcommand{\ubbA}{\underline{\mathbb{A}}}
\newcommand{\uSi}{\underline{\Sigma}}

\newcommand{\tildex}{\tilde{x}}
\newcommand{\tildey}{\tilde{y}}
\newcommand{\tildes}{\tilde{s}}
\newcommand{\hatx}{\hat{x}}
\newcommand{\hatW}{\hat{W}}

\newcommand{\hats}{\hat{s}}
\newcommand{\hatG}{\widehat{\mathcal{G}}}
\newcommand{\hatr}{\hat{r}}
\newcommand{\s}{\sigma}
\newcommand{\Si}{\Sigma}

\newcommand{\calE}{{\mathcal{E}}}
\newcommand{\calF}{{\mathcal{F}}}
\newcommand{\calG}{{\mathcal{G}}}

\newcommand{\calM}{{\mathcal{M}}}
\newcommand{\calN}{{\mathcal{N}}}
\newcommand{\calP}{{\mathcal{P}}}
\newcommand{\calR}{{\mathcal{R}}}
\newcommand{\calS}{{\mathcal{S}}}
\newcommand{\calB}{{\mathcal{B}}}
\newcommand{\calI}{{\mathcal{I}}}
\newcommand{\calT}{{\mathcal{T}}}

\newcommand{\calZ}{{\mathcal{Z}}}
\newcommand{\ep}{{\epsilon}}

\newcommand{\ee}{\end{equation}}
\newcommand{\be}{\begin{equation}}
\newcommand{\ba}{\begin{array}}
\newcommand{\ea}{\end{array}}

\let \proof \IEEEproof
\let \endproof \IEEEendproof

\begin{document}

\title{ \Large \bf
Gaussian Channels with Feedback: Optimality, Fundamental
Limitations, and Connections of Communication, Estimation, and
Control }

\thispagestyle{empty}

\author{ \normalsize {Jialing Liu and Nicola Elia
\thanks{This research was supported by NSF under Grant ECS-0093950.  The material in
this paper was presented in part at the 43rd Annual Allerton
Conference on Communication, Control, and Computing, Monticello,
IL, September 2005.  This paper was submitted to IEEE Transactions
on Information Theory in October, 2005.}
\thanks{The authors are with the Department of Electrical and Computer Engineering, Iowa State
University, Ames, IA 50011, USA (e-mail: liujl@iastate.edu, nelia@iastate.edu).}}}

\maketitle

\begin{abstract}

Gaussian channels with memory and with noiseless feedback have
been widely studied in the information theory literature. However,
a coding scheme to achieve the feedback capacity is not available.
In this paper, a coding scheme is proposed to achieve the feedback
capacity for Gaussian channels. The coding scheme essentially
implements the celebrated Kalman filter algorithm, and is
equivalent to an estimation system over the same channel without
feedback.  It reveals that the achievable information rate of the
feedback communication system can be alternatively given by the
decay rate of the Cramer-Rao bound of the associated estimation
system. Thus, combined with the control theoretic
characterizations of feedback communication (proposed by Elia),
this implies that the fundamental limitations in feedback
communication, estimation, and control coincide.  This leads to a
unifying perspective that integrates information, estimation, and
control.   We also establish the optimality of the Kalman
filtering in the sense of information transmission, a supplement
to the optimality of Kalman filtering in the sense of information
processing proposed by Mitter and Newton.  In addition, the
proposed coding scheme generalizes the Schalkwijk-Kailath codes
and reduces the coding complexity and coding delay.  The
construction of the coding scheme amounts to solving a
finite-dimensional optimization problem.  A simplification to the
optimal stationary input distribution developed by Yang, Kavcic,
and Tatikonda is also obtained. The results are verified in a
numerical example.

\end{abstract}



\section{Introduction} \label{sec:intro}

Communication systems in which the transmitters have access to
noiseless feedback of channel outputs have been widely studied. As
one of the most important case, the single-input single-output
frequency-selective Gaussian channels with feedback have attracted
considerable attention; see
\cite{kailath1,kailath2,omura68,butman-1969,butman-1976,cover-pombra-1989,
ozarow90:2,yanagi92,ordentlich,
feder:isit04,tati:capI,kavcic_it04,sahai:phd,elia_c5,kim04,kim_allerton05}
and references therein for the capacity computation and coding
scheme design for these channels. In particular,
\cite{kailath1,kailath2} proposed ingenious feedback codes (called
the Schalkwijk-Kailath codes, in short the SK codes) for additive
white Gaussian noise (AWGN) channels, which achieve the asymptotic
feedback capacity (i.e. the infinite-horizon feedback capacity,
denoted $C_\infty$) and greatly reduce the coding complexity and
coding delay. \cite{butman-1969,butman-1976,ozarow90:2} presented
the extensions of the SK codes to Gaussian feedback channels with
memory and obtained tight capacity bounds.

\cite{cover-pombra-1989} presented a rather general coding
structure (called the Cover-Pombra structure, in short the CP
structure) to achieve the finite-horizon feedback capacity
(denoted $C_T$, where the horizon spans from time epoch 0 to time
epoch $T$) for Gaussian channels with memory; however, it involves
prohibitive computation complexity as the coding length $(T+1)$
increases.  By exploiting the special properties of a
moving-average Gaussian channel with feedback, \cite{ordentlich}
discovered the finite rankness of the innovations in the CP
structure, which reduces the computation complexity.
\cite{feder:isit04} reformulated the CP structure along this
direction, and obtained an SK-based coding scheme to achieve $C_T$
with reduced computation complexity.   Also along the line of
\cite{ordentlich}, \cite{kim04} studied a first-order
moving-average Gaussian channel with feedback, found the
closed-form expression for $C_\infty$, and obtained an SK-based
coding scheme to achieve $C_\infty$.

\cite{tati:capI} provided a thorough study of feedback capacity;
extended the notion of directed information proposed
in~\cite{massey:dmi} and proved that its supremum is the feedback
capacity; reformulated the problem of computing $C_T$ as a
stochastic control optimization problem; and proposed a dynamic
programming based solution. This idea was further explored in
\cite{kavcic_it04}, which uncovered the Markov property of the
optimal input distributions for Gaussian channels with memory and
eventually reduced the finite-horizon stochastic control
optimization problem to a manageable size. Moreover, under a
\emph{stationarity conjecture} that $C_\infty$ equals the
stationary capacity (the maximum information rate over all
\emph{stationary} input distributions, denoted $C^s$), $C_\infty$
is given by the solution of a finite dimensional optimization
problem. This is the first computationally efficient
\footnote{Here we do not mean that their optimization problem is
convex. In fact the computation complexity for $C_{fb,T}$ is
$O(T+1)$, and for $C_{fb,\infty}$ the complexity is determined
mainly by the channel order, which does not involve prohibitive
computation if the channel order is not too high.} method to
calculate $C^s$ or $C_T$ for general Gaussian channels. The
stationary conjecture has been recently confirmed, namely
$C^s=C_\infty$, and $C_\infty$ is achievable using a (an
asymptotically) stationary input distribution
\cite{kim_allerton05}.

\cite{omura68} proposed a view of regarding the optimal
communication over an AWGN channel with feedback as a control
problem.  \cite{sahai:phd} investigated the problem of tracking
unstable sources over a channel and introduced the notion of
\emph{anytime capacity} to capture the fundamental limitations in
that problem, which reveals intimate connections between
communication and control and brings new insights to feedback
communication problems. Furthermore, \cite{elia_c5} established
the equivalence between feedback communication and feedback
stabilization over Gaussian channels with memory, showed that the
achievable transmission rate is given by the Bode sensitivity
integral of the associated control system, and presented an
optimization problem based on robust control to compute lower
bounds of $C^s$. \cite{elia_c5} also extended the SK codes to
achieve these lower bounds, and the coding schemes have an
interpretation of tracking unstable sources over Gaussian
channels.

For Gaussian networks with feedback, tight capacity bounds can be
found in \cite{ozarow84:mac,kramer02,elia_c5}.  For time-selective
fading channels with AWGN and with feedback, an SK-based coding
scheme utilizing the channel fading information was constructed in
\cite{liu:markov} to achieve the ergodic capacity.

As we can see, it remains an open problem to build a coding scheme
with reasonable complexity to achieve $C_\infty$ for a Gaussian
channel with memory; note that no practical codes have been found
based on the optimal signalling strategy in \cite{kavcic_it04}. In
this paper, we propose a coding scheme for frequency-selective
Gaussian channels with output feedback. This coding scheme
achieves $C_\infty$, the asymptotic feedback capacity of the
channel; utilizes the Kalman filter algorithm; simplifies the
coding processes; and shortens the coding delay. The optimal
coding structure is essentially a finite-dimensional linear
time-invariant (FDLTI) system, is also an extension of the SK
codes, and leads to a further simplification of the optimal
stationary signalling strategy in \cite{kavcic_it04}. The
construction of the coding system amounts to solving a
finite-dimensional optimization problem. Our solution holds for
AWGN channels with intersymbol interference (ISI) where the ISI is
model as a stable and minimum-phase FDLTI system; through the
equivalence shown in \cite{tati:capI,kavcic_it04}, this channel is
equivalent to a colored Gaussian channel with rational noise power
spectrums and without ISI.  Note that the rationalness assumption
is widely used and not too restrictive, since any power spectrum
can be arbitrarily approximated by rational ones.

In deriving our optimal coding design in infinite-horizon, we
first present finite-horizon analysis (which is closely related to
the CP structure) of the feedback communication problem, and then
let the horizon length tend to infinity and obtain our optimal
coding design which achieves $C_\infty$. More specifically, in our
finite-horizon analysis, we establish the necessity of the Kalman
filter: The Kalman filter is not only a device to provide
sufficient statistics (which was shown in \cite{kavcic_it04}), but
also a device to ensure the power efficiency and to recover the
message optimally.  This also leads to a refinement of the CP
structure, applicable for generic Gaussian channels. Additionally,
the presence of the Kalman filter in our coding scheme reveals the
intrinsic connections among feedback communication, estimation,
and control.  In particular, we show that the feedback
communication problem over a Gaussian channel is essentially an
optimal estimation problem, and the achievable rate of the
feedback communication system is alternatively given by the decay
rate of the Cramer-Rao bound (CRB) for the associated estimation
system. Invoking the Bode sensitivity characterization of the
achievable rate \cite{elia_c5}, we conclude that the fundamental
limitations in feedback communication, estimation, and control
coincide. We then extend the horizon to infinity and characterize
the steady-state of the feedback communication problem. We finally
show that our optimal scheme achieves $C_\infty$.

We also remark that the necessity of the Kalman filter in the
optimal coding scheme is not surprising, given various indications
of the essential role of Kalman filtering (or minimum mean-squared
error (MMSE) estimators; or minimum-energy control, its control
theory equivalence; or the sum-product algorithm, its
generalization) in optimal communication designs.  See e.g.
\cite{kschischang01,forney03,kavcic_it03,kavcic_it04,elia_c5,mitter:kalman}.
The study of the Kalman filter in the feedback communication
problem along the line of \cite{mitter:kalman} may shed important
insights on optimal communication problems and is under current
investigation.

One main insight gained in this study is that, the perspective of
unifying information, estimation, and control, three fundamental
concepts, facilitates our development of the optimal feedback
communication design.  Though the connections between any two of
the three concepts have been investigated or are under
investigation, a joint study explicitly addressing all three is
not available. Our study provides the first example that the
connections among the three can be explored and utilized, to the
best of our knowledge.  In addition to helping us to achieve the
optimality in the feedback communication problem,  this new
perspective establishes the optimality of the Kalman filtering in
the sense of information transmission, a supplement to the
optimality of Kalman filtering in the sense of information
processing proposed by Mitter and Newton \cite{mitter:kalman}. It
also leads to a new formula connecting the mutual information in
the feedback communication system and MMSE in the associated
estimation problem, a supplement to a fundamental relation between
mutual information and MMSE proposed by Guo, Shamai, and Verdu
\cite{guo:it05}. We anticipate that this new perspective may help
us to study more general feedback communication problems in future
investigations, such as multiuser feedback communications.

This paper is organized as follows.  In Section \ref{sec:pre}, we
introduce the channel models. The problem formulation is given in
Section \ref{sec:prob_sol}, followed by the problem solution, i.e.
the optimal coding scheme and the coding theorem. In Section
\ref{sec:kf}, we prove the necessity of the Kalman filter in
generating the optimal feedback.  In Section \ref{sec:dual}, we
provide the connections of the feedback communication problem to
an estimation problem and a control problem, and express the
maximum achievable rate in terms of estimation theory quantities
and control theory quantities.  In Section \ref{sec:achieve}, we
show that our coding scheme is capacity-achieving.  Section
\ref{sec:num} provides a numerical example. Finally we conclude
the paper and discuss future research directions.

\textbf{Notations:} We represent time indices by subscripts, such as $y_t$.  We denote
by $y^T$ the vector $\{y_0,y_{1},$ $\cdots,$ $y_T\}$, and $\{y_t\}$ the sequence
$\{y_t\}_{t=0}^\infty$.  We assume that the starting time of all processes is 0,
consistent with the convention in dynamical systems but different from the information
theory literature. We use $h(X)$ for the differential entropy of the random variable
$X$.  For a random vector $y^T$, we denote its covariance matrix as $K_y^{(T)}$.  For
a stationary process $\{y_t\}$, we denote its power spectrum as $\calS_y (e^{j 2\pi
\theta})$.  We denote $\calT_{xy}(z)$ as the transfer function from $x$ to $y$.  We
denote ``defined to be" as ``$:=$".  We use $(A,B,C,D)$ to represent system
\be \left\{ \ba{lll} x_{t+1} &=& A x_t + B u_t \\
y_t &=& C x_t + D u_t. \ea \right. \ee

\section{Channel model} \label{sec:pre}

In this section, we briefly describe two Gaussian channel models, namely the colored
Gaussian noise channel without ISI and white Gaussian noise channel with ISI.

\subsection{Colored Gaussian noise channel without ISI} \label{subsec:color}

\begin{figure}[h!]
\center \subfigure[ ]{{\scalebox{.5}{\includegraphics{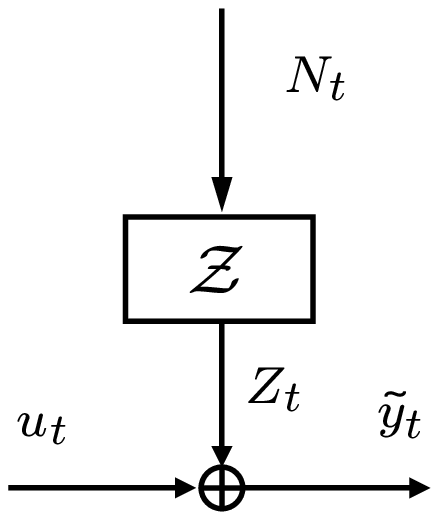}}}} \hspace{0in}
\subfigure[ ]{{\scalebox{.5}{\includegraphics{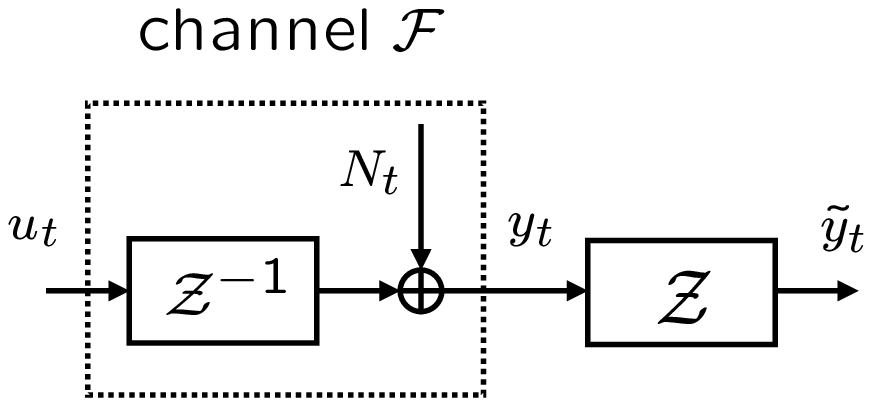}}}} \hspace{0in} \subfigure[
]{{\scalebox{.5}{\includegraphics{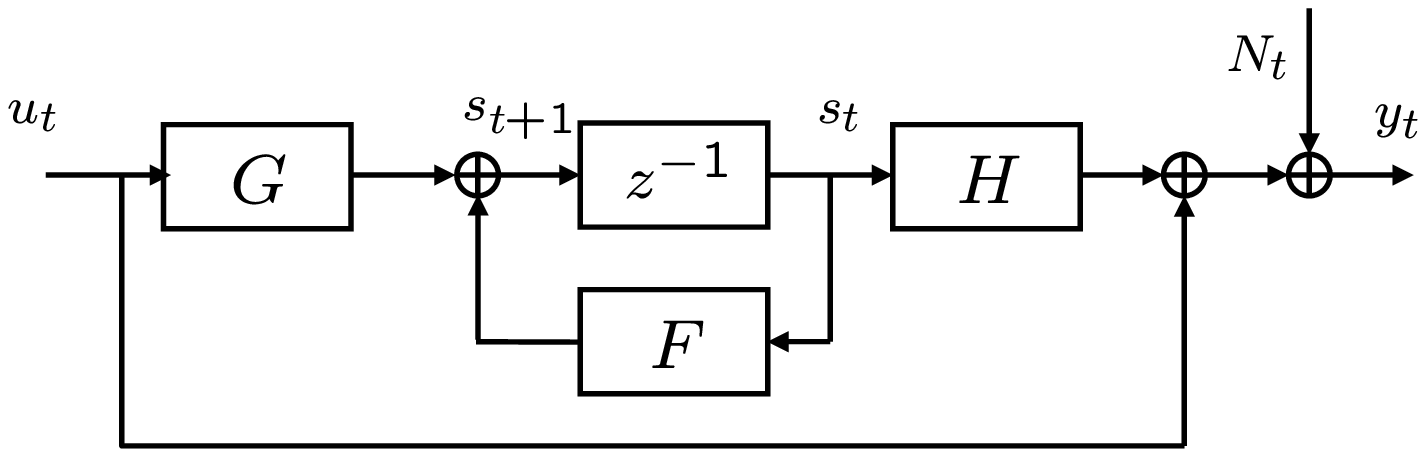}}}} \caption{ (a) A colored Gaussian noise
channel without ISI.  (b) The equivalent ISI channel with AWGN. (c) State-space
realization of channel $\calF$.} \label{fig:isicolor}
\end{figure}

Fig. \ref{fig:isicolor} (a) shows a colored Gaussian noise channel without ISI.  At
time $t$, this discrete-time channel is described as
\be \tilde{y}_t = u_t + Z_t, \;\textnormal{ for } t=0,1,\cdots, \label{chan:color} \ee
where $u_t$ is the channel input, $Z_t$ is the channel noise, and $\tilde{y}_t$ is the
channel output.  We make the following assumptions: The colored noise $\{Z_t\}$ is the
output of a finite-dimensional stable and minimum-phase linear time-invariant (LTI)
system $\calZ(z)$ driven by a white Gaussian process $\{N_t\}$ of zero mean and unit
variance, and $\calZ(z)$ is at initial rest.  For any block size (i.e. coding length)
of $(T+1)$, we may equivalently generate $Z^T$ by
\be Z^T = \calZ_T N^T, \label{noise:color}\ee
where $\calZ_T$ is a $(T+1)\times(T+1)$ lower-triangular Toeplitz matrix of the
impulse response of $\calZ(z)$.  We may abuse the notation $\calZ$ for both $\calZ(z)$
and $\calZ_T$ if no confusion arises.  As a consequence, $\{Z_t\}$ is asymptotically
stationary. \footnote{The difference between a stationarity assumption and an
\emph{asymptotic} stationarity assumption may result from different starting points of
the process: If starting from $t=-\infty$, $\{Z_t\}$ is \emph{stationary}; instead if
starting from $t=0$ as we are assuming here, $\{Z_t\}$ is \emph{asymptotically
stationary}.  They result in exactly the same steady-state analysis of the feedback
communication problem.}

Note that there is no loss of generality in assuming that $\calZ(z)$ is stable and
minimum-phase (cf. Chapter 11, \cite{papoulis}), implying that the initial condition
of $\calZ(z)$ generates no effect on the steady-state.   Thus we made the initial rest
assumption since we mainly focus on the steady-state characterization.

\subsection{White Gaussian channel with ISI} \label{subsec:isi}

The above colored Gaussian channel induces a new channel, namely a white Gaussian
channel with ISI, under a further assumption that $\calZ(\infty) \neq 0$ (i.e. $\calZ$
is proper but non-strictly proper).  More precisely, notice that from
(\ref{chan:color}) and (\ref{noise:color}), we have
\be \tilde{y}^T = \calZ_T (\calZ_T^{-1} u^T + N^T ), \ee
which we identify as a stable and minimum-phase ISI channel with AWGN $\{N_t\}$, see
Fig. \ref{fig:isicolor} (b).   Here $\calZ^{-1}(z)$ is also at initial rest.  For any
fixed $u^T$ and $N^T$, (a) and (b) generate the same channel output $\tilde{y}^T$.
\footnote{More rigorously, the mappings from $(u,N)$ to $\tilde{y}$ are
$T$-equivalent. For a discussion about systems representations and equivalence between
different representations, see Appendix \ref{appsec:equiv}.} Note that $\calZ_T^{-1}$
is the matrix inverse of $\calZ_T$, equal to the lower-triangular Toeplitz matrix of
impulse response of $\calZ^{-1}(z)$.

The initial rest assumption on $\calZ^{-1}$ can be imposed in practice equivalently
by, first driving the initial condition of the ISI channel to any desired value (known
to the receiver) before a transmission, and then removing the response due to that
initial condition at the receiver.  Such an assumption is also used in
\cite{kavcic_it04,tati:capI}. We further assume for simplicity that $\calZ(\infty)=1$;
for cases where $g:=\calZ(\infty) \neq 1$, we can normalize $\calZ(z)$ by scaling it
by $1/g$. Hence, $\calZ_T$ is a lower triangular Toeplitz matrix with diagonal
elements all equal to 1 (and thus is invertible).

We can then write the minimal state-space representation of $\calZ^{-1}$ as
$(F,G,H,1)$, where $F \in \mathbb{R}^{m}$ is stable, $(F,G)$ is controllable, $(F,H)$
is observable, and $m$ is the \emph{dimension} or \emph{order} of $\calZ^{-1}$. Let us
denote the channel from $u$ to $y$ in Fig. \ref{fig:isicolor}(b) as $\calF$, where
\be y^T := \calZ_T^{-1} u^T + N^T = \calZ_T^{-1} \tilde{y}^T. \ee
The channel $\calF$ is described in state-space as
\be  \textnormal{channel } \calF: \left \{\ba{lll} s_{t+1}&=&F s_t + G u_t \\
y_t &=& H s_t +u_t +N_t, \ea \right.\ee
where $s_0=0$; see Fig. \ref{fig:isicolor} (c).  Notice that channel $\calF$ is not
essentially different than the channel from $u$ to $\tilde{y}$, since $\{y^t\}$ and
$\{\tilde{y}^t\}$ causally determine each other.

We concentrate on the case $m \geq 1$; the case that $m$ is 0 (i.e., $\calF$ is an
AWGN channel) was solved in \cite{kailath1,kailath2}.

\section{Problem formulation in steady-state and the solution}
\label{sec:prob_sol}

Before formulating the steady-state communication problem, we
distinguish among the three scenarios: Finite-horizon (i.e. finite
coding length), infinite-horizon (i.e. infinite coding length),
and steady state.  Finite-horizon problems often have
time-dependent (i.e. time-varying) and horizon-dependent solutions
(similar to finite-horizon Kalman filtering).  The
horizon-dependence may be removed in the infinite-horizon
scenario, and furthermore, the time-dependence may be removed in
the steady-state scenario. If we find the (stationary,
time-invariant) steady-state solution (which by
\cite{kim_allerton05} is also the infinite-horizon solution), we
can truncate it and employ the truncation to the practical problem
in finite-horizon provided that the horizon is large enough. This
truncated solution would greatly simplify the implementation while
having a performance sufficiently close to finite-horizon
optimality.

\subsection{Problem formulation}

For a Gaussian channel with feedback, the channel input may take the form
\be u_t=\gamma_t u^{t-1} + \eta_t y^{t-1} + \xi_t \label{u1}\ee
for any $\gamma_t \in \bbR^{1 \times t}$, $\eta_t \in \bbR^{1
\times t}$, and zero-mean Gaussian random variable $ \xi_t \in
\bbR$ which is independent of $u^{t-1}$ and $y^{t-1}$ (cf.
\cite{tati:capI,kavcic_it04}).  Therefore, the channel inputs are
allowed to depend on the channel outputs in a strictly causal
manner. Our objective in this paper is to \emph{design
encoder/decoder to achieve the asymptotic feedback capacity},
given by
\be \ba{lcl} C_\infty:= C_\infty(\calP) := & \disp \sup_{\{u_t\}
\textrm{ \small stationary,(\ref{u1})}}& \disp \lim _{T
\rightarrow \infty} \frac{1}{T+1}I(u^T
\rightarrow y^T) \\
&\disp  ^{s.t. \: P_\infty:= \lim _{T \rightarrow \infty}  \E u^T{}' u^T /(T+1) \leq
\mathcal{P} } & \ea \label{opt:imax}\ee
where $\calP>0$ is the power budget and $I(u^T \rightarrow y^T)$
is the directed information from $u^T$ to $y^T$ (cf.
\cite{tati:capI}).   For more details about $C_\infty$, refer to
\cite{kavcic_it04,kim_allerton05} and Section \ref{subsec:GM} in
this paper.

The problem of solving $C_\infty$ may be equivalently formulated
as minimizing the average channel input power while keeping the
information rate bounded from below, namely for $\calR>0$,
\be \ba{lcl} P_\infty(\calR) :=  &\displaystyle \inf_{\{u_t\}
\textrm{  \small stationary,(\ref{u1})}}  &   \disp \lim _{T
\rightarrow \infty} \frac{1}{T+1} \E u^T{}' u^T.
\\ &\disp  ^{s.t. \:  \lim _{T \rightarrow \infty} I(u^T
\rightarrow y^T) /(T+1) \geq \calR  } & \ea \label{opt:imax_pmin}\ee
Therefore $P_\infty(\calR)$ is the inverse function of
$C_\infty(\calP)$, i.e., $C_\infty(P_\infty(\calR))=\calR$.

\textbf{Approach:} Our approach to solve the steady-state
communication problem is to investigate the finite-horizon problem
first, and then let the horizon increase to infinity, which leads
to a unified treatment of infinite-horizon and finite-horizon.
Other approaches not pursued in this paper are also possible, such
as applying the idea in \cite{elia_c5} to the optimal signalling
strategy in \cite{kavcic_it04}, though they generate results not
as rich as the present approach does.

\subsection{The coding scheme}

The rest of this section presents the solution to the above
problem.  In this subsection, we introduce an encoder/decoder
structure and explain how to choose the parameters to ensure the
optimality, and then describe the encoding/decoding process, that
is, how we assign the message to be transmitted, and how we
recover the message. In the next subsection, we present the coding
theorem which states that our encoding/decoding structure with the
chosen parameters achieves $C_\infty$. The proof of the theorem
will be developed in Sections \ref{sec:kf} to \ref{sec:achieve}.

\textbf{The encoder/decoder structure}

\begin{figure}[h!]
\begin{center}
{{\includegraphics[width=\textwidth]{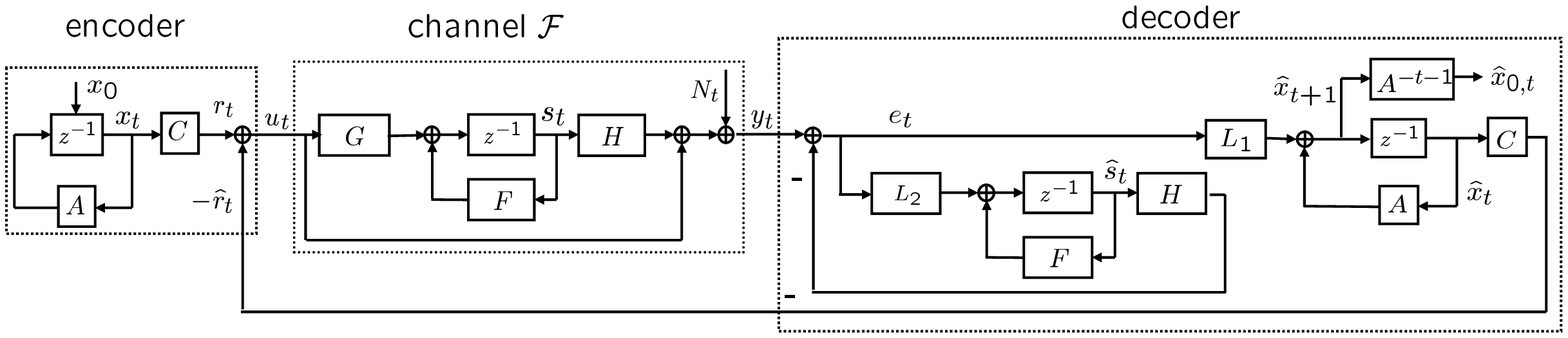}}} \caption{The encoder/decoder structure
for $\calF$.} \label{fig:liu1}
\end{center}
\end{figure}

In state-space, the encoder and decoder are described as
\be  \textnormal{Encoder:} \left \{ \ba{lll} x_{t+1} &=& Ax_t\\
r_t &=& C x_t \\
u_t &=& r_t - \hatr_t \ea \right. \label{enc}\ee
and
\be \textnormal{Decoder:} \left\{ \ba{lll} \hats_{t+1} &=& F \hats_{t} + L_{2} e_t\\
e_t &=& y_t - H \hats_t \\
\hatx_{t+1} &=& A \hatx_{t} + L_{1} e_t\\
\hatr_t &=& C \hatx_t\\
\hat{x}_{0,t} &=& A^{-t-1} \hatx_{t+1}, \ea \right.\label{dec}\ee
where $\hats_0=0$, $\hatx_0=0$, $A \in \bbR^{(n+1)\times (n+1)}$, $C \in \bbR^{1
\times (n+1)}$, $L_{1} \in \bbR ^{n+1}$, and $L_{2} \in \bbR^m$.  We call $(n+1)$ the
\emph{encoder dimension}, $x_t$ the \emph{encoder state}, and $\hatx_{0,t}$ the
\emph{decoder estimate}. See Fig. \ref{fig:liu1} for the block diagram. Observe that
$-\hatr_t$ is the feedback from the decoder based on the channel output $y^{t-1}$, and
thus $u_t$ depends on $y^{t-1}$ but not $y_t$.   It further follows that $ -\hatr^t =
\calG^*_t y^t $ for some strictly lower triangular Toeplitz matrix $\calG^*_t$.  Here
$A,C,u_t$, etc. depend on $n$, but we do not specify the dependence explicitly to
simplify notations.

\textbf{Optimal choice of parameters}

Fix a desired rate $\calR$.  Let $DI:=2^\calR$ and $n :=m-1$ (recalling that $m$ is
the channel dimension), and solve the optimization problem
\begin{equation}
\begin{array}{lcl}
[\bfa_f^{opt},\Si^{opt}]:= & \displaystyle{\textnormal{arg} \inf_{\bfa_f \in \bbR^{n}}}  &  \bbD \Sigma \bbD',  \\
     & ^ {s.t. \: \Sigma = \bbA \Sigma \bbA' - \bbA \Sigma \bbC'  \bbC \Sigma \bbA'/(\bbC \Sigma
\bbC ' +1)}
\end{array} \label{opt:dare}\vspace{-10pt}
\end{equation}
where
\be  \bbA :=\left[ \ba{c|c} A & 0 \cr \hline GC & F \ea \right], \bbC :=\left[
\matrix{ C & H } \right],  \bbD :=\left[ \matrix{ C & 0 } \right] ,A :=\left[ \ba{c|c}
0_{n \times 1} & I_n \cr \hline \pm DI & \bfa_f \ea \right]  , C :=\left[ \ba{cc} 1&
0_{1 \times n} \ea \right] .  \ee
Note that we need to solve (\ref{opt:dare}) twice (one for $+DI$ in $A$ and one for
$-DI$ in $A$), and choose the optimal solution as the one with the smaller objective
function value. Then we form the optimal $A^{opt}$ based on $\bfa_f^{opt}$, and let
$(n^*+1)$ be the number of unstable eigenvalues in $A^{opt}$, where $n^*\geq 0$.

Now let $n:=n^*$, solve (\ref{opt:dare}) again, and obtain a new $\bfa_f^{opt}$ and
$\Si^{opt}$.  Then form $A^{opt}$, let $A^*=A^{opt}$, $\Si^*=\Si^{opt}$,
$C^*:=[1,0_{1\times n^*}]$, and form $\bbA^*$,$\bbC^*$, and $\bbD^*$.  Let
\be L^* :=[L_1^*{}',L_2^*{}']':=\frac{\bbA^* \Si^* \bbC^*{}' }{\bbC^* \Sigma^*
\bbC^*{} ' +1}. \ee
As we will show, $(A^*,C^*)$ is observable, and $A^*$ has exactly $(n^*+1)$ unstable
eigenvalues.

We assign the encoder/decoder parameters to the scheme built in Fig. \ref{fig:liu1} by
letting
\be n:=n^*,A:=A^*, C:=C^*, L_1:=L_1^*,L_2:=L_2^*.\ee
We then drive the initial condition $s_0$ of channel $\calF$ to
zero.  Now we are ready to communicate at a rate $\calR$ using
power $P_\infty(\calR)= \bbD^* \Sigma^* \bbD^*{}'$. \footnote{We
see from (\ref{opt:dare}) that for \emph{any} channel $\calF$, a
simple upper bound of the function $P_\infty(\calR)$ is given by
$\min \{ (2^{2\calR}-1)(\calZ(2^{\calR}))^2
,(2^{2\calR}-1)(\calZ(-2^{\calR}))^2 \}$, obtained by using one
unstable eigenvalue in $A$. }

\textbf{Encoding/Decoding process}

\subsubsection{Transmission of analog source}

The designed communication system can transmit either an analog source or a digital
message.  In the former case, we assume that the encoder wishes to convey a Gaussian
random vector through the channel and the decoder wishes to learn the random vector,
which is a rate-distortion problem (or successive refinement problem, see e.g.
\cite{tati:phd,sahai:phd,gastpar04}). The coding process is as follows. Assume that
the to-be-conveyed message $W$ is distributed as $\calN(0,I_{n^*+1})$ (noting that any
non-degenerate $(n^*+1)$-variate Gaussian vector $W$ can be transformed into this
form). Assume that the coding length is $(T+1)$.  To encode, let $x_{0}:=W$. Then run
the system till time epoch $T$, obtaining $\hatx_{0,t}$, $t = 0,1,\cdots,T$. To
decode, let $\hatW_t:= \hatx_{0,t}$ for $t = 0,1,\cdots,T$.

The quantities of interest include the squared-error distortion,
defined as
\be \MSE (\hatW_{t}):=\E (W-\hatW_{t})(W-\hatW_{t})'. \ee
It will become clear that $\MSE (\hatW_{t})$ can be pre-computed before the
transmission, and thus the coding length can be determined \emph{a priori} to ensure a
desired distortion level.

\subsubsection{Transmission of digital message}

To transmit digital messages over the communication system, let us first fix $\ep>0$
small enough and the coding length $(T+1)$ large enough.  Let
\be \Si_x^* : = [I_{n^*+1},0] \Si^* [I_{n^*+1},0]' .\ee
Assume that the matrix $(A^*{}')^{-T-1} \Si_x^* (A^*)^{-T-1}$ has an eigenvalue
decomposition as
\be (A^*{}')^{-T-1} \Si_x^* (A^*)^{-T-1} = E_T \Lambda_T E_T' , \ee
where $E_T=[e^{(1)},\cdots,e^{(n^*+1)}]$ is an orthonormal matrix and $\Lambda_T$ is a
positive diagonal matrix.  Let $\sigma_{T,i}$ be the square root of the $(i,i)$th element
of $\Lambda_T$. Let $\calB \in \mathbb{R}^{n^*+1}$ be the unit hypercube spanned by
columns of $E_T$, that is,
\be \calB=\left\{ \left. \sum_{i=0}^{n^*} \alpha^{(i)} e^{(i)} \right| \alpha^{(i)}
\in [-\frac{1}{2},\frac{1}{2}],i=0,\cdots,n^* \right\}. \ee
Next we partition the $i$th side of $\calB$ into $( \sigma_{T,i} ) ^{-(1-\epsilon)}$
segments. This induces a partition of $\calB$ into $M_T$ sub-hypercubes, where
\be \ba{lll}M_{T} & =&  \disp \prod _{i=0} ^{n^*} \left(\sigma_{T,i}\right)
^{-(1-\epsilon)}
\\
&=&  \disp \left[\det \left( (A^*{}')^{-T-1} \Si^* (A^*)^{-T-1} \right)\right]
^{-\frac{1-\epsilon} {2} }. \ea \ee
We then map the sub-hypercube centers to a set of $M_T$ equally likely messages. The
above procedure is known to both the transmitter and receiver \textit{a priori}.

Suppose now we wish to transmit the message represented by the center $W$. To encode,
let $x_{0}:=W$.  Then run the system till time epoch $T$.  To decode, we map
$\hatx_{0,T}$ into the closest sub-hypercube center and obtain the decoded message
$\hatW_T$.  We declare an error if $\hatW_T \neq W$, and call a (an asymptotic) rate
\be R := \lim _{T \rightarrow \infty} \frac{1}{T+1} \log M_T \ee
achievable if the probability of error $PE_T$ vanishes as $T$ tends to infinity.  We
remark that this coding process is the one used in \cite{elia_c5} for Gaussian
channels with memory, which was an extension of the SK codes.  In fact, the original
SK coding scheme can be rewritten in a Kalman filter form, and hence it essentially
implements the Kalman filtering algorithm.  We also remark that, similar to the analog
transmission case, the coding length $(T+1)$ can be pre-determined.

As we have seen, the encoder/decoder design and the
encoding/decoding process can be done rather easily.  The
computation complexity for encoding/decoding grows as $O(T+1)$.
Also interestingly, the encoder may be viewed as a control system,
and the decoder may be viewed as an estimation system, as pointed
out by Sanjoy Mitter and in \cite{sahai:phd,sekharII}.

\subsection{Coding theorem}

\begin{theorem} \label{th:main}
Construct the encoder/decoder shown in Fig. \ref{fig:liu1} using $n^*$, $A^*$, $C^*$,
$L_1^*$, and $L_2^*$.  Then under the power constraint $\E u^2 \leq \calP$,

i) The coding scheme transmits an analog source $W \sim
\calN(0,I_{n^*+1})$ from the encoder to the decoder at rate
$C_\infty(\calP)$, with MSE distortion $\MSE(\hat{W}_T)$ achieving
the optimal asymptotic rate-distortion tradeoff given by
\be  R= \lim _{T \rightarrow \infty} \frac{1}{2(T+1)} \log
\frac{1} {\det \MSE(\hat{W}_T)} .\ee

ii) The coding scheme can transmit digital message from the
encoder to the decoder at a rate arbitrarily close to
$C_\infty(\calP)$, with $PE_T$ decays to zero doubly
exponentially.
\end{theorem}

The proof of the theorem will be developed in the subsequent four sections.  In
Section \ref{sec:kf}, we consider a general coding structure in finite-horizon which
may be viewed as a generalization of our optimal coding structure.  We show that this
general structure essentially contains a Kalman filter. The presence of the Kalman
filter links the feedback communication problem to an estimation problem and a control
problem, and hence we rewrite the information rate in terms of estimation theory
quantities and control theory quantities; see Section \ref{sec:dual}.  Sections
\ref{sec:kf} and \ref{sec:dual} are focused on finite-horizon. In Section
\ref{sec:asym}, we extend the horizon to infinity and characterize the steady-state
behavior.  Then in Section \ref{sec:achieve}, we show that our optimal encoder/decoder
design is actually the solution to the steady-state communication problem.

\section{Necessity of Kalman filter for optimal coding} \label{sec:kf}

In this section, we consider a finite-horizon coding structure
that includes our optimal design in Section \ref{sec:prob_sol} as
a special case. This general structure is useful since: 1)
searching over all possible parameters in the general structure
achieves $C_\infty$, that is, there is no loss of generality or
optimality to focus on this structure only; 2) we can show that to
ensure power efficiency (to be explained), the general structure
necessarily contains a Kalman filter.  The general coding
structure is in fact a variation of the CP structure (see Appendix
\ref{appsub:cp_relation}), and hence our Kalman filter
characterization leads to a refinement of the CP structure.

\subsection{A general coding structure} \label{sub:generalstructure}

\begin{figure}[h!]
\begin{center}
{\scalebox{.5}{\includegraphics{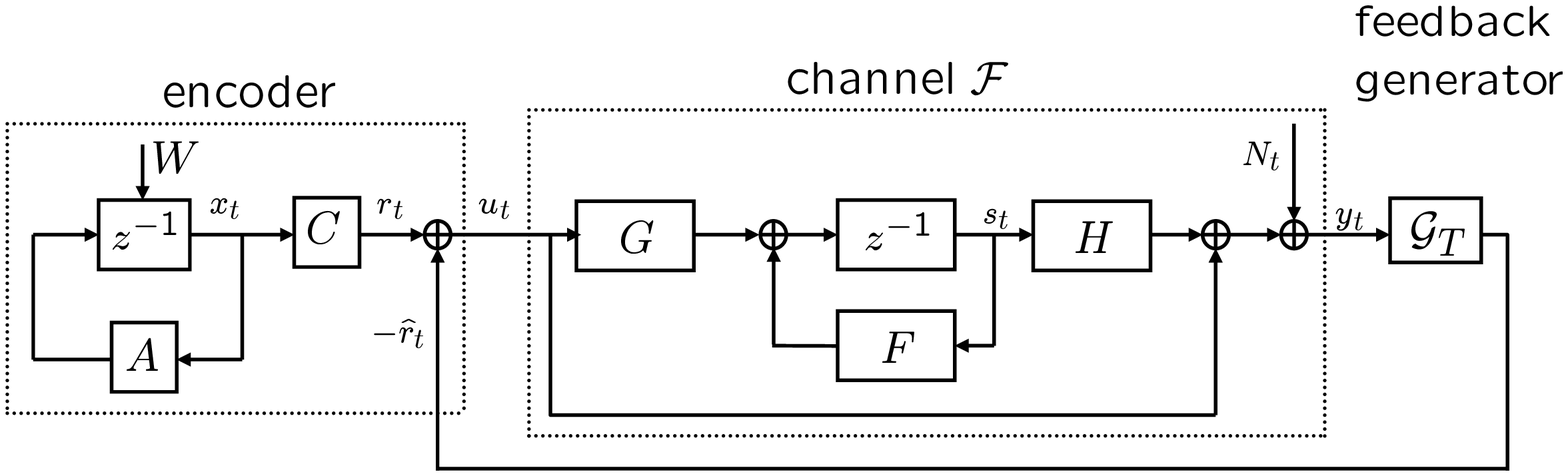}}} \caption{A general coding structure for
channel $\calF$.} \label{fig:liu2}
\end{center}
\end{figure}

Fig. \ref{fig:liu2} illustrates the general coding structure, including the encoder
and the \emph{feedback generator}, a portion of the decoder. Below, we fix the time
horizon to be $\{0,1,\cdots,T\}$ and describe the coding structure.

\textbf{Encoder:} The encoder follows the dynamics (\ref{enc}).  We assume that the
encoder dimension $(n+1)$ satisfies $0 \leq n \leq T$, $W \sim \calN(0,I_{n+1})$, $A
\in \bbR^{(n+1)\times (n+1)}$, $C \in \bbR^{1 \times (n+1)}$, $(A,C)$ is observable,
and none of the eigenvalues of $A$ are on the unit circle or at the locations of the
eigenvalues of $F$.  We then let
\be \ba{lllll} \Gamma_n(A,C)&:=& \Gamma_n&:=&[C',A'C',\cdots,A^{n}{}' C']'  \\
\Gamma(A,C)&:=&\Gamma&:=& [C',A'C',\cdots,A^{T}{}' C']' \\
K_r^{(T)}(A,C)&:=&K_r^{(T)}&:=&\E r^T r^T{}'. \ea \ee
Therefore, $\Gamma_n$ is the observability matrix for $(A,C)$ and is invertible,
$\Gamma$ has rank $(n+1)$, $r^T=\Gamma W$, and $K_r^{(T)} = \Gamma \Gamma'$.

\textbf{Feedback generator:} The feedback signal $-\hatr_t$ is generated through the
\emph{feedback generator} $\calG_T$, i.e.
\be -\hatr^T= \calG_T y^T.  \label{fbgen} \ee
We assume that $\calG_T \in \bbR^{(T+1)\times (T+1)}$ is a strictly lower triangular
matrix. Clearly, the optimal encoder/decoder can be viewed as a special case of the
general structure. Throughout the paper, the above assumptions on the encoder/decoder
are always assumed unless otherwise specified.  For future use purpose, we compute the
channel output as
\be y^T = (I-\calZ_T^{-1} \calG_T) ^{-1} (\calZ_T^{-1} r^T + N^T). \ee

\begin{definition}
Consider the general coding structure shown in Fig. \ref{fig:liu2}.  Define
\be \ba{lll} \displaystyle C_{T,n}:=  C_{T,n}(\calP)&:=& \displaystyle \sup _{A \in
\bbR^{(n+1)\times (n+1)},C,\calG_T} \frac{1}{T+1}I(W;y^T) \\
& &  ^{s.t. \: \E u^T{}' u^T/(T+1) \leq \mathcal{P} } \ea \vspace{-10pt} \ee
and define its inverse function as $P_{T,n}(\calR)$.
\end{definition}

In other words, $C_{T,n}$ is the finite-horizon information
capacity for a \emph{fixed transmitter dimension} $n$.  It holds
that $C_{n,n}=C_n$ and hence $\lim _{n \rightarrow \infty}
C_{n,n}=C_\infty$ (see Lemma \ref{lemma:ctn} and Appendix
\ref{appsub:c_inf}). Moreover, as we will show, $C_\infty$ can be
achieved using this structure.

\subsection{The presence of Kalman filter} \label{sub:nec4kf}

We first compute the mutual information in the general coding structure.

\begin{prop} \label{prop:I}
Consider the general coding structure in Fig. \ref{fig:liu2}. Fix any $0 \leq n \leq
T$, and fix any $A,C$, and $\calG_T$.  Then it holds that
\be \ba{lll} I(W;y^T)&=& I(r^T;y^T) \\
& = & I(u^T \rightarrow y^T) \\
& = &  \frac{1}{2} \log \det K_y^{(T)}\\
&=& \frac{1}{2} \log \det (I+ \calZ_T^{-1} K_r^{(T)}
\calZ_T^{-1'}) \\
&=& \frac{1}{2} \log \det (I+ \calZ_T^{-1} \Gamma \Gamma' \calZ_T^{-1'}), \ea \ee
which is \emph{independent} of $\calG_T$.
\end{prop}

\proof
\be  \ba{lll}  I(W;y^T) & = &h(y^T) - h(y^T|W) \\
&=& h(y^T) - h \left((I-\calZ_T^{-1} \calG_T) ^{-1} (\calZ_T^{-1} r^T + N^T) | W \right) \\
&\eqa& \frac{1}{2} \log \det (2 \pi e K_y^{(T)}) - h(N^T)  \\
&\eqb& I(u^T \rightarrow y^T) \\
&=&\frac{1}{2} \log \det K_y^{(T)} \\
&= & \frac{1}{2} \log \det (I+ \calZ_T^{-1} K_r^{(T)} \calZ_T^{-1'}) , \ea \ee
where (a) is due to $r^T=\Gamma W$, $\det (AB)=\det A \det B$, and
$\det(I-\calZ_T^{-1} \calG_T) ^{-1}=1$; and (b) follows from \cite{elia_c5}. \endproof

Proposition \ref{prop:I} implies that $I(W;y^T)$ is independent of the feedback
generator $\calG_T$, and dependent only on $K_r^{(T)}$ or equivalently on $(A,C)$.
Thus, fixed $(A,C)$ implies \emph{fixed information rate}, and hence the optimal
feedback generator has to be chosen to minimize the average channel input power, which
turns out to contain a Kalman filter.  Note that the counterpart of this proposition
in infinite-horizon was proven in \cite{elia_c5}. Now we can define, for a fixed
$(A,C)$, the information rate across the channel to be
\be R_T(A,C):= \frac{I(W;y^T)}{T+1} .\ee

The optimal feedback generator for a given $(A,C)$ is found in the next proposition.

\begin{figure}[h!]
\center \subfigure[ ]{\scalebox{.45}{\includegraphics{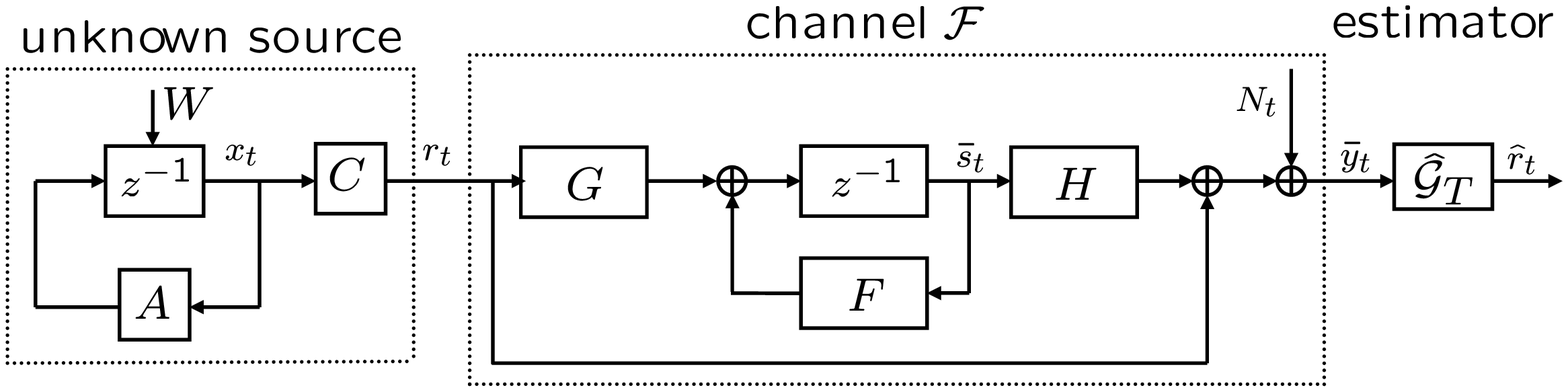}}}  \subfigure[
]{\scalebox{.45}{\includegraphics{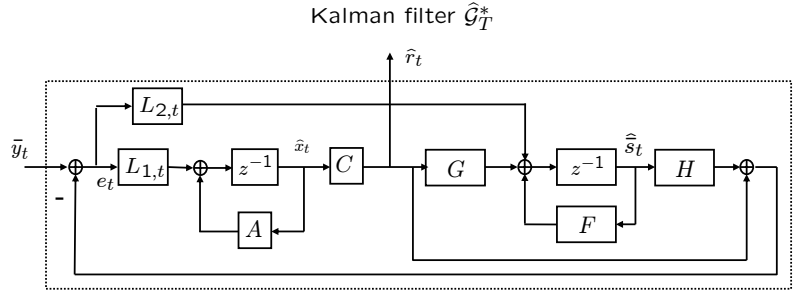}}} \subfigure[
]{{\scalebox{.45}{\includegraphics{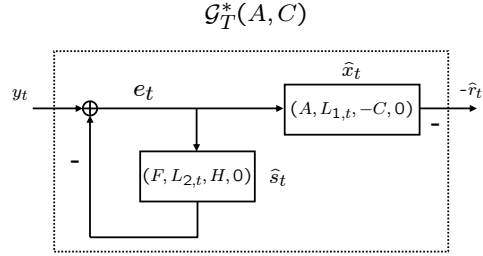}}}} \caption{ (a) An estimation problem over
channel $\mathcal{F}$.  (b) The Kalman filter $\hatG_T^*(A,C)$.  (c) The Kalman filter
based feedback generator $\calG_T^*(A,C)$.  Here $(A,L_{1,t},-C,0)$ with $\hatx_t$
denotes a state-space representation with $\hatx_t$ being its state at time $t$, and
$\hatx_0$ being 0; see (\ref{def:bba}) and (\ref{ric:recur}) for $L_{1,t}$ and
$L_{2,t}$.} \label{fig:estim}
\end{figure}

\begin{prop} \label{prop:kf} Consider the general coding structure in Fig.
\ref{fig:liu2}. Fix any $0 \leq n \leq T$.  Then

i) \be \ba{llcllcl}P_{T,n}(\calR) &= &   \displaystyle
\inf_{A,C,\calG_T:=\calG_T^*(A,C)}
 & \displaystyle \frac{1}{T+1}\E u^T{}' u^T \\
 & & ^{s.t. \: R_T(A,C) \geq  \calR} & \ea
\label{opt:calG1}\ee
where $\calG_T^*(A,C)$ is the \emph{optimal feedback generator} for a given $(A,C)$,
defined as
\be \calG_T^*(A,C) :=    \arg \displaystyle \inf _{(A,C) \textnormal{ \small
fixed},\calG_T} \displaystyle \frac{1}{T+1} \E u^T{}' u^T .\label{opt:calG2} \ee

ii) The optimal feedback generator $\calG_T^*(A,C)$ is given by
\be \calG_T^*(A,C) = - \hatG_T^*(A,C) (I- \calZ_T^{-1} \hatG_T^*(A,C)) ^{-1} ,
\label{eq:calGhatG}\ee
where $\hatG_T^*(A,C)$ is the strictly causal MMSE estimator (Kalman filter) of $r^T$
given the noisy observation $ \bar{y}^T := \calZ_T^{-1} r^T+N^T $, i.e.,
\be \ba{lll} \hatG_T^*(A,C) &:=&   \displaystyle \arg \inf_{\hatG_T \in
\bbR^{(T+1)\times(T+1)}}  \displaystyle \frac{1}{T+1} \E (r^T-\hatG_T
\bar{y}^T)(r^T-\hatG_T \bar{y}^T)', \ea \label{opt:calG3}  \ee
where $\hatG_T$ is strictly lower triangular. See Fig. \ref{fig:estim} (a) for the
associated estimation problem, (b) for the Kalman filter $\hatG_T^*(A,C)$, and (c) for
the optimal feedback generator $\calG_T^*(A,C)$.

\end{prop}

\begin{remark} \rm Proposition \ref{prop:kf} reveals that, the minimization of channel input
power in a feedback communication problem is equivalent to the minimization of MSE in
an estimation problem.  This equivalence yields a complete characterization (in terms
of the Kalman filter) of optimal feedback generator $\calG_T^*(A,C)$ for a given
$(A,C)$. Since our general coding structure is a variation of the CP structure, this
proposition leads to the Kalman filter based characterization of the CP structure and
hence is an improvement of the Cover-Pombra formulation; see Appendix
\ref{appsub:cp_relation}.
\end{remark}

\begin{remark} \rm Proposition \ref{prop:kf} i) implies that we may reformulate the problem of $C_{T,n}$
(or $P_{T,n}$) as a two-step problem: In step 1, we fix $(A,C)$,
i.e. fixing the rate, and minimize the input power by searching
over $\calG$; and in step 2, we search over all possible $(A,C)$
subject to the rate constraint.  The \emph{role of the feedback
generator $\calG$} for any fixed $(A,C)$ is to minimize the input
power. Then ii) solves the optimal feedback generator
$\calG_T^*(A,C)$ by considering the \emph{equivalent} optimal
estimation problem in Fig. \ref{fig:estim} (a) whose solution is
the Kalman filter. Notice that the Kalman filter can also give us
the optimal estimate of the message $W$.  Hence, the Kalman filter
leads to both \emph{power efficiency} and \emph{the best estimate
of the message}. The power efficiency is ensured by the one-step
prediction operation of the Kalman filtering, and the optimal
recovery of message is ensured by the smoothing operation of the
Kalman filtering; therefore, we obtain the optimality of Kalman
filtering in the information transmission sense. We finally note
that the necessity of the Kalman filter is not surprising given
the previous indications in
\cite{kailath2,butman-1976,sahai:phd,tati:capI,mitter:kalman},
etc.
\end{remark}

\proof  i) Notice that for any fixed $(A,C)$, $R_T (A,C)$ is fixed.  Then from the
definition of $P_{T,n}(\calR)$, we have
\be \ba{llcll}P_{T,n}(\calR) &= &   \displaystyle \inf_{A,C,\calG_T}
 & \displaystyle \frac{1}{T+1}\E u^T{}' u^T \\
 & & ^{s.t. \: R_T(A,C) \geq  \calR} &  \\
 &= &   \displaystyle \inf_{A,C}
 & \disp \inf_{(A,C) \textnormal{ \small fixed},\calG_T}
 & \displaystyle \frac{1}{T+1}\E u^T{}' u^T \\
 & & ^{s.t. \: R_T(A,C) \geq  \calR} . &   \ea
\ee
Then i) follows from the definition of $\calG^*_T(A,C)$.

ii) Note that for the general coding structure, it holds that
\be u^T=r^T +(-\hatr^T) =r^T+\calG_T y^T.\label{eq:input}\ee
Then, letting
\be \hatG_T:= -\calG_T(I-\calZ_T^{-1} \calG_T)^{-1} \ee
and $\bar{y}^T:= \calZ_T^{-1} r^T+N^T$, we have $\calG_T y^T= -\hatG_T \bar{y}^T$.
Therefore,
\be \ba{lll} \calG_T^*(A,C) &=&   \displaystyle \arg \inf_{\calG_T} \displaystyle
\frac{1}{T+1} \E (r^T+\calG_T
y^T)(r^T+\calG_T y^T)' \\
&=&   \displaystyle \arg \inf_{\hatG_T} \displaystyle  \frac{1}{T+1} \E (r^T-\hatG_T
\bar{y}^T)(r^T-\hatG_T \bar{y}^T)'. \ea   \ee
The last equality implies that the optimal solution $\hatG_T^*$ is the strictly causal
MMSE estimator (with one-step prediction) of $r^T$ given $\bar{y}_T$; notice that
$\hatG_T$ is strictly lower triangular. It is well known that such an estimator can be
implemented recursively in state-space as a Kalman filter (cf.
\cite{book:kay1,kailath:book}).  Finally, from the relation between $\calG_T$ and
$\hatG_T$, we obtain (\ref{eq:calGhatG}).  The state-space representation of
$\calG_T^*(A,C)$ needs only a straightforward computation, as shown in Appendix
\ref{appsec:equiv}.
\endproof

We remark that it is possible to derive a dynamic programming based solution
(\cite{tati:capI}) to compute $C_{T,n}$, and if we further employ the Markov property
in \cite{kavcic_it04} and the above Kalman filter based characterization, we would
reach a solution with complexity $O(T)$ for computing $C_{T,n}$ and $C_T$.  However,
we do not pursue along this line in this paper since it is beyond the main scope of
this paper.

\section{Feedback rate, CRB, and Bode integral} \label{sec:dual}

We have shown that in the general coding structure, to ensure power efficiency for a
fixed $(A,C)$, we need to design a Kalman-filter based feedback generator. The Kalman
filter immediately links the feedback communication problem to estimation and control
problems.  In this section, we present a \emph{unified representation} for the general
coding structure (with $\calG$ being chosen as $\calG^*(A,C)$), its estimation theory
counterpart, and its control theory counterpart. Then we will establish connections
among the information theory quantities, estimation theory quantities, and control
theory quantities.

\subsection{Unified representation of feedback coding system, Kalman filter,
and minimum-energy control} \label{subsec:dual}

In this subsection, we will present the dynamics for the estimation problem and the
general coding structure, then show that they are governed by one set of equations,
which may also be viewed as a control system.

\textbf{The estimation system}

The estimation system in Fig. \ref{fig:estim} consists of three parts: the unknown
source $r^T$ to be estimated or tracked, the channel $\calF$ (without output
feedback), and the estimator which we choose as the Kalman filter $\hatG^*$; we assume
that $(A,C)$ is fixed and known to the estimator.  The system is described in
state-space as
\be \textnormal{estimation system:}
\left\{
\ba{ll}
        \ba{lll}    x_{t+1} &=& A x_t \\
                        r_t &=& Cx_t
        \ea  & \left. \ba{l}\\  \ea \right\} \textnormal{unknown source}      \\
        \ba{lll}    \bar{s}_{t+1}  &=& F \bar{s}_t + G r_t \\
                        \bar{y}_t &=& H \bar{s}_t + r_t + N_t
        \ea  & \left. \ba{l}\\  \ea \right\} \textnormal{channel }\calF \\
        \ba{lll}    \hatx_{t+1} &=& A \hatx_t + L_{1,t} e_t \\
                        \hatr_t &=& C \hatx_t\\
                        \hat{\bar{s}}_{t+1} &=&  F \hat{\bar{s}}_{t} + G \hatr_t + L_{2,t} e_t\\
                        e_t &=& \bar{y}_t - H \hat{\bar{s}}_{t}  -\hatr_t
        \ea  & \left. \ba{l}\\ \\ \\ \\  \ea \right\} \textnormal{Kalman filter }\hatG^*(A,C)
\ea
\right. \label{dyn:est}
\ee
with $x_0=W$, $\bar{s}_0=\hat{\bar{s}}_0=0$, and $\hat{x}_0=0$.
Here $L_{1,t} \in \bbR ^{n+1}$ and $L_{2,t} \in \bbR^m$ are the
time-varying Kalman filter gains specified in (\ref{eq:Lt}).

\textbf{The general coding structure with the optimal feedback generator}

The optimal feedback generator for a given $(A,C)$ is solved in (\ref{eq:calGhatG}),
see Fig. \ref{fig:estim} (c) for its structure.  We can then obtain the minimal
state-space representation of $\calG^*_T(A,C)$, and describe the general coding
structure with $\calG^*_T(A,C)$ as
\be \textnormal{general coding structure:}
    \left\{ \ba{ll}
    \ba{lll} x_{t+1} &=& A x_t \\
            r_t &=& Cx_t\\
            u_t &=& r_t - \hatr_t
    \ea  & \left. \ba{l}\\ \\ \ea \right\} \textnormal{encoder}           \\
    \ba{lll}  s_{t+1} &=& Fs_t + G u_t \\
                y_t &=& Hs_t + u_t +N_t
        \ea  & \left. \ba{l}\\  \ea \right\} \textnormal{channel }\calF \\
    \ba{lll}  \hats_{t+1} &=& F \hats_{t} + L_{2,t} e_t \\
                e_t &=& y_t - H \hats_t \\
                \hatx_{t+1} &=& A \hatx_{t} + L_{1,t} e_t \\
                -\hatr_t &=& -C \hatx_t
        \ea  & \left. \ba{l}\\ \\ \\   \ea \right\} \textnormal{optimal feedback generator }\calG^*(A,C)
                \ea \right. \label{dyn:coding}\ee
with $x_0=W$, $s_0=\hats_0=0$, and $\hatx_0=0$. See Appendix \ref{appsec:equiv} for
the derivation of the minimal state-space representation of $\calG^*_T(A,C)$.  It can
be easily shown that $r_t$, $\hatr_t$, $e_t$, $x_t$, and $\hatx_t$ in (\ref{dyn:est})
and (\ref{dyn:coding}) are equal, respectively, and it holds that
\be s_t - \hats_t = \bar{s}_t - \hat{\bar{s}}_t. \ee

\textbf{The unified representation}

\begin{figure}[h!]
\begin{center}
{\scalebox{.5}{\includegraphics{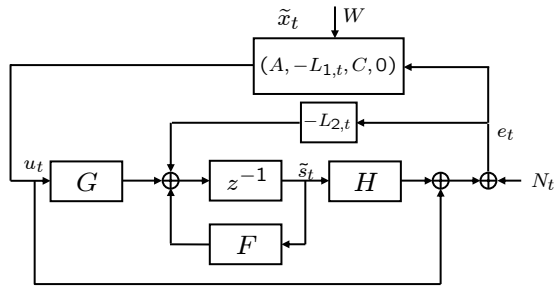}}} \caption{The block diagram for the
minimum-energy control system. Here the block $(A,-L_{1,t},C,0)$ with $\tilde{x}_t$
denotes the state-space representation with $\tilde{x}_t$ and $W$ being its state at
time $t$ and at time 0.} \label{fig:menergy}
\end{center}
\end{figure}

Define
\be \ba{lll} \tilde{x}_t &:=& x_t - \hatx_t \\
\tilde{s}_t &: =& s_t - \hats_t = \bar{s}_t - \hat{\bar{s}}_t\\
\bbX_t &:=&\left[ \matrix{\tilde{x}_{t} \cr \tilde{s}_{t}} \right] \\
\bbX_0 &:=&\left[ \matrix{W \cr 0} \right]\\
\bbA &:=&\left[ \ba{c|c} A & 0 \\ \hline GC & F  \ea \right] \\
\bbC &:=&\left[ \matrix{ C & H } \right] \\
\bbD &:=&\left[ \matrix{ C & 0 } \right] \\
L_t &:=&\left[ \matrix{L_{1,t} \cr L_{2,t}} \right] . \ea \label{def:bba}\ee
Note that $\bbX_t$ is the estimation error for $[x_t',s_t']'$.  Substituting
(\ref{def:bba}) to (\ref{dyn:est}) and (\ref{dyn:coding}), we obtain that both systems
become
\be \textnormal{control system:}\left\{ \ba{lll} \bbX_{t+1} &=& (\bbA - L_t \bbC) \bbX_t - L_t N_t=\bbA \bbX_t-L_t e_t\\
 e_t &=& \bbC \bbX_t+N_t \\
 u_t &=& \bbD \bbX_t; \ea \right. \label{dyn}\ee
see Fig. \ref{fig:menergy} for its block diagram.  It is a control system where we
want to minimize the power of $u$ by appropriately choosing $L_t$.  This is a
\emph{minimum energy control} problem, which is useful for us to characterize the
steady-state solution and it is equivalent to the Kalman filtering problem (see
\cite{mincontrol:book}).

The signal $e_t$ in (\ref{dyn}) is called the \emph{Kalman filter innovation} or
\emph{innovation} \footnote{The innovation defined here is different from the
innovation defined in \cite{cover-pombra-1989} or \cite{kavcic_it04}.}, which plays a
significant role in Kalman filtering. One fact is that $\{e_t\}$ is a white process,
that is, its covariance matrix $K_e^{(T)}$ is a diagonal matrix. Another fact is that
$e^T$ and $y^T$ determine each other causally, and we can easily verify that
$h(e^T)=h(y^T)$ and $\det K_y^{(T)}= \det K_e^{(T)}$.  We remark that (\ref{dyn}) is
the \emph{innovations representation} of the Kalman filter (cf. \cite{kailath:book}).

For each $t$, the optimal $L_t$ is determined as
\be L_t : = \left[ \ba{l}L_{1,t} \\ L_{2,t} \ea \right] :=
\frac{\bbA \Si_t \bbC'}{ K_{e,t}}, \label{eq:Lt}\ee
where $\Si_t:= \E \bbX_t \bbX_t'$, $K_{e,t} := \E (e_t)^2  = \bbC \Si_t \bbC'+1$, and
the error covariance matrix $\Si_t$ satisfies the Riccati recursion
\be \Si_{t+1} = \bbA \Sigma_t \bbA' - \frac{\bbA \Sigma_t \bbC' \bbC \Si_t \bbA'
}{\bbC \Sigma_t \bbC ' +1} \label{ric:recur} \ee
with initial condition
\be \Si_0 :=\left[ \matrix{ I_{n+1} & 0 \cr 0 & 0 } \right] , \label{ric:init}\ee
This completes the description of the optimal feedback generator for a given
$(A,C)$.

The meaning of a unified expression for three different systems (\ref{dyn:est}),
(\ref{dyn:coding}), and (\ref{dyn}) is that the first two are actually two different
non-minimal realizations of the third.  The input-output mappings from $N^T$ to $e^T$
in the three systems are $T$-equivalent (see Appendix \ref{appsub:equiv}). Thus we say
that the three problems, the optimal estimation problem, the optimal feedback
generator problem, and the minimum-energy control problem, are \emph{equivalent} in
the sense that, if any one of the problems is solved, then the other two are solved.
Since the estimation problem and the control problem are well studied, the equivalence
facilitates our study of the communication problem. Particularly, the formulation
(\ref{dyn}) yields alternative expressions for the mutual information and average
channel input power in the feedback communication problem, as we see in the next
subsection.

\begin{figure}[h!]
\center{{\scalebox{.5}{\includegraphics{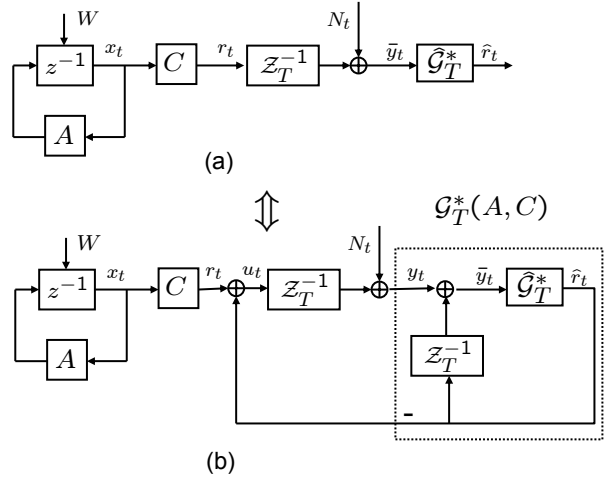}}}} \caption{ Relation between the
estimation problem (a) and the communication problem (b).} \label{fig:dual}
\end{figure}

We further illustrate the relation of the estimation system and the communication
system in Fig. \ref{fig:dual}: (b) is obtained from (a) by subtracting $\hatr_t$ from
the channel input and adding $\calZ_T^{-1} \hatr_t$ back to the channel output, which
does not affect the input, state, and output of $\hatG_T^*$.  It is clearly seen from
the block diagram manipulations that the minimization of channel input power in
feedback communication problem becomes the minimization of MSE in the estimation
problem.

\subsection{Mutual information in terms of Fisher information and
CRB} \label{subsec:crb}

\begin{prop} \label{prop:dual}

For any fixed $0\leq n \leq T$ and $(A,C)$, it holds that

i) \be \ba{lll}I(W;y^T)& =& \disp \frac{1}{2} \log \det K_e^{(T)} = \frac{1}{2} \sum _{t=0}^T \log  K_{e,t}\\
& =&  \disp \frac{1}{2} \sum _{t=0}^T \log  (\bbC \Si_t \bbC'+1 )\\
& =&  \disp \frac{1}{2} \log \det \MMSE_{W,T}^{-1} \\
& =&  \disp \frac{1}{2} \log \det \calI_{W,T} \\
&=& \disp \frac{1}{2} \log \det \textnormal{CRB}_{W,T} ^{-1}; \ea \ee

ii) \be \ba{lll}P_{T,n}(A,C) &=&  \disp \frac{1}{T+1} \sum _{t=0} ^T \bbD
\Si_t \bbD' \\
&=&  \disp \frac{1}{T+1} \trace (\CMMSE_{r,T})\\
&=&  \disp \frac{1}{T+1} \sum _{t=0}^T CA^t \MMSE_{W,t} A^t{}'C' , \ea \ee
where $\MMSE_{W,T}$ is the minimum MSE of $W$, $\CMMSE_{r,T}$ is the causal minimum
MSE of $r^T$, $\calI_{W,T}$ is the Bayesian Fisher information matrix of $W$ for the
estimation system (\ref{dyn:est}), and $\textnormal{CRB}_{W,T}$ is the Bayesian CRB of
$W$~\cite{vantrees}.

\end{prop}

\begin{remark} \rm This proposition connects the mutual information to the innovations process and to the
Fisher information, (minimum) MSE, and CRB of the associated
estimation problem. As a consequence, the finite-horizon feedback
capacity $C_{T,n}$ is then linked to the smallest possible
Bayesian CRB, i.e. the smallest possible estimation error
covariance, and thus the fundamental limitation in information
theory is linked to the fundamental limitation in estimation
theory.  It is also interesting to notice that the Fisher
information, an estimation quantity, indeed has an information
theoretic interpretation as its name suggests.  Besides, the link
between the mutual information and the MMSE provides a supplement
to the fundamental relation discovered in \cite{guo:it05}; the
connections between our result and that in \cite{guo:it05} is
under current investigation.
\end{remark}

\proof i) First we simply notice that $h(y^T)=h(e^T)$, and $K_{e,t} = \bbC \Si_t
\bbC'+1$.  Next, to find MMSE of $W$, note that in Fig. \ref{fig:estim} (a)
\be \bar{y}^T = \calZ_T^{-1} \Gamma W+N^T \ee
and that $W \sim \calN(0,I)$, $N^T \sim \calN(0,I)$.  Thus, by \cite{book:kay1} we
have
\be \MMSE_{W,t} = (I + \Gamma' \calZ_T^{-1} {}' \calZ_T^{-1} \Gamma)^{-1} , \ee
yielding
\be \ba{lll} \det \MMSE_{W,t} &=& \det (I + \calZ_T^{-1} \Gamma
\Gamma' \calZ_T^{-1} {}' )^{-1} \\
&=&\det (I + \calZ_T^{-1} K_r^{(T,n)}\calZ_T^{-1} {}' )^{-1} . \ea \ee
Besides, from Section 2.4 in \cite{vantrees} we can directly compute the FIM of $W$ to
be $(I + \Gamma' \calZ_T^{-1} {}' \calZ_T^{-1} \Gamma)$.  Then i) follows from
Proposition \ref{prop:I} and (\ref{dyn}).

ii) Since $u_t=\bbD \bbX_t = C \tilde{x}_t =r_t-\hatr_t$ and $\E \tilde{x}_t
\tilde{x}_t' =A^t \MMSE_{W,t} A^t{}'$, we have $\E (u_t)^2=\bbD \Si_t \bbD'= C \E
\tilde{x}_t \tilde{x}_t' C'= \E (r_t-\hatr_t)^2$, and then ii) follows. \endproof

\subsection{Necessary condition for optimality} \label{subsec:kim}

Before we turn to the infinite-horizon analysis, we show in this subsection that our
general coding structure together with the optimal feedback generator satisfies a
``necessary condition for optimality" discussed in \cite{kim04}.  The condition says
that, the channel input $u_t$ needs to be orthogonal to the past channel outputs
$y^{t-1}$. This is intuitive since to ensure fastest transmission, the transmitter
should not transmit any information that the receiver has obtained, thus the
transmitter wants to remove any correlation of $y^{t-1}$ in $u_t$ (to this aim, the
transmitter has to access the channel outputs through feedback).

\begin{prop} \label{prop:kim}
In system (\ref{dyn:coding}), for any $0 \leq \tau < t$, it holds that $\E u_t
y_\tau=0$.
\end{prop}

\proof See Appendix \ref{appsub:kim}.  \endproof

\section{Asymptotic behavior of the system} \label{sec:asym}

By far we have completed our analysis in finite-horizon.  We have shown that the
optimal design of encoder and decoder must contain a Kalman filter, and connected the
feedback communication problem to an estimation problem and a control problem. Below,
we consider the steady-state communication problem, by studying the limiting behavior
($T$ going to infinity) of the finite-horizon solution while fixing the encoder
dimension to be $(n+1)$.

\subsection{Convergence to steady-state}  \label{subsec:asym}

The time-varying Kalman filter in (\ref{dyn}) converges to a steady-state, namely
(\ref{dyn}) is \emph{stabilized} in closed-loop, $u_t$, $e_t$, and $y_t$ will converge
to steady-state distributions, and $\Si_t$, $L_t$, $\calG^*_t(A,C)$, $\hatG_t^*$, and
$K_{e,t}$ will converge to their steady-state values.  That is, asymptotically
(\ref{dyn}) becomes an LTI system
\be \textnormal{steady-state:}\left\{ \ba{lll} \bbX_{t+1}
&=& (\bbA - L \bbC) \bbX_t - L N_t=\bbA \bbX_t-L e_t\\
 e_t &=& \bbC \bbX_t+N_t \\
 u_t &=& \bbD \bbX_t, \ea \right. \label{dyn:steady}\ee
where
\be L : = \frac{\bbA \Si \bbC'}{ K_{e}}, \ee
$K_{e} = \bbC \Si \bbC'+1$, and $\Si$ is the unique stabilizing solution to the
Riccati equation
\be \Si = \bbA \Sigma \bbA' - \frac{\bbA \Sigma \bbC' \bbC \Si \bbA' }{\bbC \Sigma
\bbC ' +1}. \label{ric:eq} \ee

This LTI system is easy to analyze (e.g., it allows transfer function based study) and
to implement.  For instance, the minimum-energy control (cf. \cite{mincontrol:book})
of an LTI system claims that the transfer function from $N$ to $e$ is an
\emph{all-pass} function in the form of
\be \calT_{Ne} (z) = \prod _{i=0}^k \frac{z-a_i}{z-a_i^{-1}} \label{allpass}\ee
where $a_0,\cdots,a_k$ are the unstable eigenvalues of $A$ or $\bbA$ (noting that $F$
is stable).  Note that this is consistent with the whiteness of innovations process
$\{e_t\}$.

The existence of steady-state of the Kalman filter is proven in the following
proposition. Notice that (\ref{dyn}) is a \emph{singular} Kalman filter since it has
no process noise; the convergence of such a problem was established in
\cite{gallivan_riccati05}.

\begin{prop} \label{prop:dare} Consider the Riccati recursion (\ref{ric:recur}) and
the system (\ref{dyn}).

i) Starting from the initial condition given in (\ref{ric:init}), the Riccati
recursion (\ref{ric:recur}) generates a sequence $\{\Si_t\}$ that converges to
$\Si_\infty$ with rank $(n+1)$, the unique stabilizing solution to the Riccati
equation (\ref{ric:eq}).

ii) The time-varying system (\ref{dyn}) converges to the unique steady-state as given
in (\ref{dyn:steady}).
\end{prop}

\proof See Appendix \ref{appsub:convg}. \endproof

\subsection{Steady-state quantities}

Now fix $(A,C)$ and let the horizon $T$ in the general coding structure go to
infinity.  Let $\mathcal{H}(e)$ be the entropy rate of $\{e_t\}$, $DI(A):=\prod
_{i=0}^k |a_i|$ be the \emph{degree of instability} of $A$, and $S (e^{j 2\pi
\theta})$ be the spectrum of the sensitivity function of system (\ref{dyn:steady})
(cf. \cite{elia_c5}). Then the limiting result of Proposition \ref{prop:dual} is
summarized in the next proposition.

\begin{prop} \label{prop:steady}

Consider the general coding structure in Fig. \ref{fig:liu2}. For any $ n \geq 0$ and
$(A,C)$,

i) The asymptotic information rate is given by
\be \ba{lll}  R_{\infty,n}(A,C) &: =& \displaystyle \lim  _{T
\rightarrow \infty} \frac{1}{T+1} I(W;y^T) \\
&=&  \displaystyle \mathcal{H}(e) - \frac{1}{2} \log 2 \pi e \\
&=& \displaystyle \log DI(A) \\
&=& \displaystyle \displaystyle \int_{-\frac{1}{2}} ^{\frac{1}{2}}
\log S (e^{j 2\pi \theta}) d \theta \\
& =& \displaystyle \frac{1}{2}  \log  (\bbC \Si \bbC'+1 )\\
 &=& \displaystyle \lim _{T \rightarrow \infty}\frac{\log \det \calI_{W,T}}{2(T+1)} \\
 &=& \displaystyle - \lim _{T \rightarrow \infty}\frac{\log \det \MSE_{W,T}}{2(T+1)} \\
&=& \displaystyle -\lim _{T \rightarrow \infty}\frac{\log  \det  \CRB_{W,T}}{2(T+1)} .
\ea \ee

ii) The average channel input power is given by
\be \ba{lll} P_{\infty,n}(A,C) & :=& \disp \lim _{T \rightarrow \infty} \frac{1}{T+1}
\E u_T
u_T'\\
&=& \displaystyle \bbD \Si \bbD' . \ea \ee

\end{prop}

\begin{remark} \rm Proposition \ref{prop:steady} links the asymptotic information rate to the entropy
rate of the innovations process, to the degree of instability and Bode sensitivity
integral (\cite{elia_c5}), to the asymptotic increasing rate of the Fisher
information, and to the asymptotic decay rate of MSE and of CRB.  Recall that the Bode
sensitivity integral is the fundamental limitation of the disturbance rejection
(control) problem, and the asymptotic decay rate of CRB is the fundamental limitation
of the recursive estimation problem.  Hence, the fundamental limitations in feedback
communication, control, and estimation coincide. \end{remark}

\begin{remark} \rm Proposition \ref{prop:steady} implies that the presence of stable eigenvalues in
$A$ does not affect the rate (see also \cite{elia_c5}). Stable eigenvalues do not
affect $P_{\infty,n}(A,C)$, either, since the initial condition response associated
with the stable eigenvalues can be tracked with zero power (i.e. zero MSE). So, we can
achieve $C_{\infty,n}$ by a sequence of purely unstable $(A,C)$, and hence the
communication problem is related to the tracking of purely unstable source over a
communication channel (\cite{sahai:phd,elia_c5}). \end{remark}

\proof Proposition \ref{prop:steady} leads to that, the limits of the results in
Proposition \ref{prop:dual} are well defined.  Then
\be \ba{lll}  R_{\infty,n}(A,C) & =& \displaystyle \lim  _{T
\rightarrow \infty} \frac{1}{2(T+1)} \sum_{t=0}^T \log K_{e,t} \\
& =& \displaystyle \lim  _{T
\rightarrow \infty} \frac{1}{2}  \log K_{e,t} \\
&=&  \displaystyle \mathcal{H}(e) -\frac{1}{2} \log 2\pi e, \ea \ee
where the second equality is due to the Cesaro mean (i.e., if $a_k$ converges to $a$,
then the average of the first $k$ terms converges to $a$ as $k$ goes to infinity), and
the last equality follows from the definition of entropy rate of a Gaussian process
(cf. \cite{cover}).

Now by (\ref{allpass}), $\{e_t\}$ has a flat power spectrum with magnitude $DI(A)^2$.
Then $R_{\infty,n}(A,C)=\log DI(A)$.  The Bode integral of sensitivity follows from
\cite{elia_c5}.  The other equalities are the direct applications of the Cesaro mean
to the results in Proposition \ref{prop:dual}.
\endproof

\section{Achievability of $C_\infty$} \label{sec:achieve}

In this section, we will prove that $C_{\infty,m-1}= C_\infty$,
leading to the optimality of our encoder/decoder design in Section
\ref{sec:prob_sol} in the mutual information sense, and then show
that our design achieves $C_\infty$ in the operational sense.

\subsection{The optimal Gauss-Markov signalling strategy and a simplification}
\label{subsec:GM}

\cite{kavcic_it04} proved that for each input in the form of (\ref{u1}), there exists
a Gauss-Markov (GM) input that yields the same directed information and same input
power. The GM input takes the form
\be u_t = d_t' \tilde{s}_{s,t} + \calE_t , \label{gm}\ee
where $d_t \in \bbR^m$ is a time-varying gain; $\{\calE_t\}$ is a
zero-mean white Gaussian process and $\calE_t$ is independent on
$N^{t-1}$, $u^{t-1}$, and $y^{t-1}$; and $\tilde{s}_{s,t}$ is
generated by a Kalman filter (noting that this Kalman filter is
different from the Kalman filter obtained in this paper)
\be \left\{ \ba{lll} \tilde{s}_{s,t}&:=&s_t - \hats_{s,t} \\
\hats_{s,t+1} &=& F \hats_{s,t} + L_{s,t} e_t \\
e_t &=& y_t - H \hats_{s,t} ,  \ea \right. \label{ykt:decoder}\ee
where $\hats_{s,0}=0$,
\be L_{s,t} : = \frac{Q_t \Si_{s,t} (H+d_t')' + K_{\calE}^{(t)} G }{1+ K_{\calE}^{(t)}
+ (H+d_t') \Si_{s,t} (H+d_t')' } ,\ee
$Q_t:=F+G d_t'$, and $\Si_{s,t}:=\E \tildes_{s,t} \tildes_{s,t}'$ is the estimation
error covariance of $s_t$, satisfying the Riccati recursion
\be   \Si_{s,t+1} =  Q_t \Sigma_{s,t} Q_t' + K_\calE^{(t)} G G'- \frac{(Q_t \Si_{s,t}
(H+d_t')' + K_\calE^{(t)}
      G)(Q_t \Si_{s,t} (H+d_t')' + K_\calE^{(t)} G)'}{1+ K_{\calE}^{(t)} +
      (H+d_t') \Si_{s,t} (H+d_t')'}
.\ee

If one lets $d_t=d$ and $K_{\calE}^{(t)}=K_{\calE}$ for all $t$,
that is, the input $\{u_t\}$ is a stationary process, then the
search over all possible $d$ and $K_{\calE}$ solves $C_\infty$,
that is,

\be \label{opt:ykt}
\begin{array}{ccl}
C_\infty(\calP) = &\displaystyle{ \max_{d \in \bbR^{m}, K_\calE
\in \bbR}} & \displaystyle \frac{1}{2} \log (1+ K_\calE + (H+d')
\Si_s (H+d')')

\end{array} \ee
subject to Riccati equation constraint and power constraint
\be \ba{lll}  \Sigma_s &= & Q \Sigma_s Q' + K_\calE GG'- \frac{(Q \Si_s (H+d')' +
K_\calE G)(Q \Si_s (H+d')' + K_\calE G)'}{1+ K_\calE + (H+d') \Si_s
(H+d')'}\\
       \calP &=& d' \Si_s d + K_\calE . \ea \label{opt:ykt_constr}\ee
We remark that \cite{kavcic_it04} was focused more on the structure of the optimal
input distribution and capacity computation, instead of designing a coding scheme; how
to encode/decode a message (rather than using a random coding argument) is not clear
from \cite{kavcic_it04}.

Now we prove that $K_{\calE}=0$, namely $\{\calE_t\}$ vanishes in steady-state.
\footnote[6]{$K_\calE=0$ was also conjectured and numerically verified by Shaohua Yang
(personal communication).}  This leads to a further simplification of the results in
\cite{kavcic_it04}.

\begin{prop} \label{prop:e0} For the GM input (\ref{gm}) to achieve $C_\infty$, it
must hold that $K_\calE=0$.
\end{prop}

\proof See Appendix \ref{appsec:e0}.  \endproof

The vanishing of $\{\calE_t\}$ in steady-state helps us to show
that, our general coding structure shown in Fig. \ref{fig:liu2}
can achieve $C_\infty$, and the encoder dimension needs not be
higher than the channel dimension, namely to achieve $C_\infty$ we
need $A$ to have at most $m$ unstable eigenvalues, as we will see
in the next subsection.

\subsection{Generality of the general coding structure; finite dimensionality of
the optimal solution} \label{subsec:general}

In this subsection, we show that the general coding structure is
sufficient to achieve mutual information $C_\infty$.  In other
words, if we search over all admissible parameters $A,C,\calG_T$
in the general coding structure, allowing $T$ to increase to
infinity and $n$ to increase to $(m-1)$, then we can obtain
$C_\infty$. Thus, there is no loss of generality and optimality to
consider only the general coding structure with encoder dimension
no greater than $m$.

\begin{definition}
Consider the general coding structure in Fig. \ref{fig:liu2}.  Let
\be \displaystyle C_{\infty,n}:= C_{\infty,n}(\calP):= \displaystyle \sup _{A \in
\bbR^{(n+1)\times (n+1)},C,\calG_\infty} \lim _{T \rightarrow \infty}
\frac{1}{T+1}I(W;y^T)  \label{def:c_infn} \ee
subject to
\be P_{\infty,n}:= \lim _{T \rightarrow \infty} \frac{1}{T+1} \E u^T{}' u^T \leq
\mathcal{P}. \label{constr:cinfn} \ee
\end{definition}

In other words, $C_{\infty,n}$ is the infinite-horizon information capacity for a
fixed transmitter dimension.  Note that $C_{\infty,n}$ exists and is finite.  To see
this, note Proposition \ref{prop:steady}, $C_{\infty,n} \leq C_\infty < \infty$, and
the fact that
\be
 C_{\infty,n}(\calP) = \sup _{A \in
\bbR^{(n+1)\times (n+1)},C,\calG^*(A,C),(\ref{constr:cinfn})} R_{\infty,n} (A,C).
\label{cinfnRinfn} \ee
The function $C_{\infty,n}(\calP)$ also induce $P_{\infty,n}(\calR)$, the ``capacity"
in terms of minimum input power subject to an information rate constraint.

\begin{prop} \label{prop:fdim} Consider the general coding structure in Fig. \ref{fig:liu2}.

i) $C_{\infty,n}$ is increasing in $n$;

ii) For channel $\calF$ with order $m \geq 1$,
$C_{\infty,n}=C_\infty$ for $n \geq m-1$.
\end{prop}

\proof See Appendix \ref{appsub:fdim}. \endproof

This proposition suggests that, to achieve $C_\infty$, we may
first fix the transmitter dimension as $(n+1)$ and let the
dynamical system run to time infinity, obtaining $C_{\infty,n}$,
and then increase $n$ to $(m-1)$.  The finite dimensionality of
the optimal solution is important since it guarantees that we can
achieve $C_\infty$ without solving an infinite-dimensional
optimization problem.

\subsection{Achieving $C_\infty$} \label{subsec:achieve}

In this subsection, we prove that our coding scheme achieves
$C_\infty$ in the information sense as  well as in the operational
sense.

\begin{prop} \label{prop:IeqImax} For the coding scheme described in Theorem
\ref{th:main}, $R_{\infty,n^*} (A^*,C^*) = C_\infty(\calP)$ and
\\$P_{\infty,n^*}$ $ (A^*,C^*)=\calP$.
\end{prop}

\proof See Appendix \ref{appsub:IeqImax}. \endproof

\begin{prop} \label{prop:main_RD} The system constructed in Theorem \ref{th:main}
transmits the analog source $W \sim \calN(0,I)$ at a rate
$C_\infty(\calP)$, with MSE distortion $D(C_\infty(\calP))$, where
$D(\cdot)$ is the distortion-rate function.
\end{prop}

\proof See Appendix \ref{appsub:main_RD}. \endproof

\begin{prop} \label{prop:main} The system constructed in Theorem \ref{th:main}
transmits a digital message $W$ from the transmitter to the
receiver at a rate arbitrarily close to $C_\infty(\calP)$ with
$PE_T$ decays doubly exponentially.
\end{prop}

\proof See Appendix \ref{appsub:main}. \endproof

Note that, the coding length needed for a pre-specified performance level can be
pre-determined since $\Si_{x,T}^*$ can be solved off-line. Besides, because the
probability of error decays doubly exponentially, it leads to much shorter coding
length than forward transmission.

\section{Numerical example} \label{sec:num}

\begin{figure}[h!]
\psfragscanon  \psfrag{alpha}{ $C_\infty$}
 \center
{{\scalebox{.38}{\includegraphics{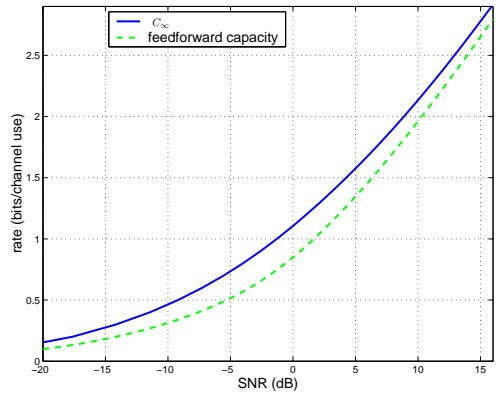}}}}
\caption{\label{fig:fbff} The asymptotic feedback capacity
$C_\infty$ and feedforward capacity for channel $\calF$ with
$\calZ^{-1}=(1+0.5 z^{-1} - 0.4 z^{-2})/(1+0.6z^{-2} -0.4 z^{-3}
)$. }
\end{figure}

Here we repeat the numerical example studied in \cite{kavcic_it04}.  Consider a
third-order channel (i.e. $m=3$) with
\be \calZ^{-1}:=\frac{1+0.5 z^{-1} - 0.4 z^{-2}}{1+0.6z^{-2} -0.4 z^{-3} } .\ee
In state-space, $\calZ^{-1}$ is described as $(F,G,H,1)$ where
\be \ba{llllll} F&=&\left[\matrix{ 0 & -0.6 & 0.4 \cr 1 & 0 & 0 \cr 0 & 1 & 0
}\right] &
\;G&=&\left[\matrix{1 \cr 0 \cr 0}\right]  \\
H&=&\left[\matrix{ 0.5 & -1 & 0.4 }\right] .&  \ea \ee
Assume the desired communication rate $\calR$ is 1 bit per channel
use.  We first solve (\ref{opt:dare}) with $n=m-1=2$, and find out
$n^*=1$.  That is, $C_\infty$ is attained when $A$ has two
unstable eigenvalues.  Then we solve (\ref{opt:dare}) again with
$n^*=1$, and obtain
\be \ba{llllll} A^*&=&\left[\matrix{ 0 & 1  \cr -2 & -0.887}\right]    \\
L^*& =& [  \matrix{-0.506&-0.225&0.573&0.092&-0.327}]'.&  \ea  \label{num:optal} \ee
This yields the optimal power $P_\infty=0.743$ (or -1.290 dB).
Similar computation generates Figure \ref{fig:fbff}, the curve of
$C_\infty$ against SNR or equivalently $\calP$.  This curve is
identical to that in~\cite{kavcic_it04}.

\begin{figure}[h!]
 \center
{{\includegraphics[width=\textwidth]{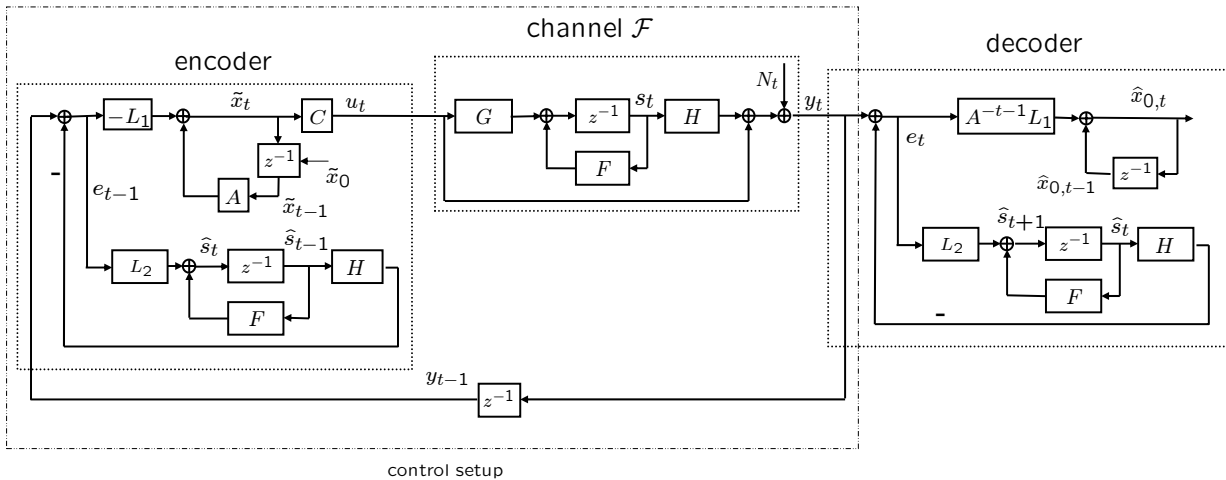}}} \caption{\label{fig:modify} The
modified feedback communication scheme. }
\end{figure}

We then use the obtained $A^*$, $C^*$, and $L^*$ to construct the optimal
communication scheme.  However, we observe that the optimal communication scheme shown
in Fig. \ref{fig:liu1} generates unbounded signals $\{r_t\}$ and $\{\hatr_t\}$ due to
the instability of $A$.  This is not desirable for the simulation purpose, though the
scheme in the form of Fig. \ref{fig:liu1} is convenient for the analysis purpose.
Here, we propose a modification of the scheme, see Fig. \ref{fig:modify}.  It is
easily verify that the system in Fig. \ref{fig:modify} is $T$-equivalent to that in
Fig. \ref{fig:liu1}.  As we indicate in Fig. \ref{fig:modify}, the loop including the
encoder, the channel, and the feedback link is indeed the control setup, which is
stabilized and hence any signal inside is bounded. \footnote{We remark that, in the
case of an AWGN channel, the modification coincides with the one studied by Gallager
(p. 480, \cite{gallager}) with minor differences.  This modification differs from the
more popular feedback communication designs in \cite{kailath1,kailath2,elia_c5};
notice that, \cite{kailath1} involves exponentially growing bandwidth, \cite{kailath2}
involves an exponentially growing parameter $\alpha^k$ where $\alpha>1$ and $k$
denotes the time index, and \cite{elia_c5} generates a feedback signal with
exponentially growing power. Thus we consider our modification more feasible for
simulation purpose.  However, this modification is not yet ``practical", mainly
because of the strong assumption on the noiseless feedback.  A more practical design
is under current investigation.}  Note that the encoder now involves $\tildex_{-1}$;
we set $\tildex_{-1}:=A^{-1}W$, leading to $\tildex_0=W$, the desired value for
$\tildex_0$.

\begin{figure}[ht!]
\psfrag{x0h1}{ \Large $\hatx_{0,t}^{(1)}$} \psfrag{x0h2}{\Large
$\hatx_{0,t}^{(2)}$}
\psfragscanon \psfrag{t}{  \LARGE \textsf{time} \LARGE$t$} \center
\subfigure[ ]{\scalebox{.492}{\includegraphics{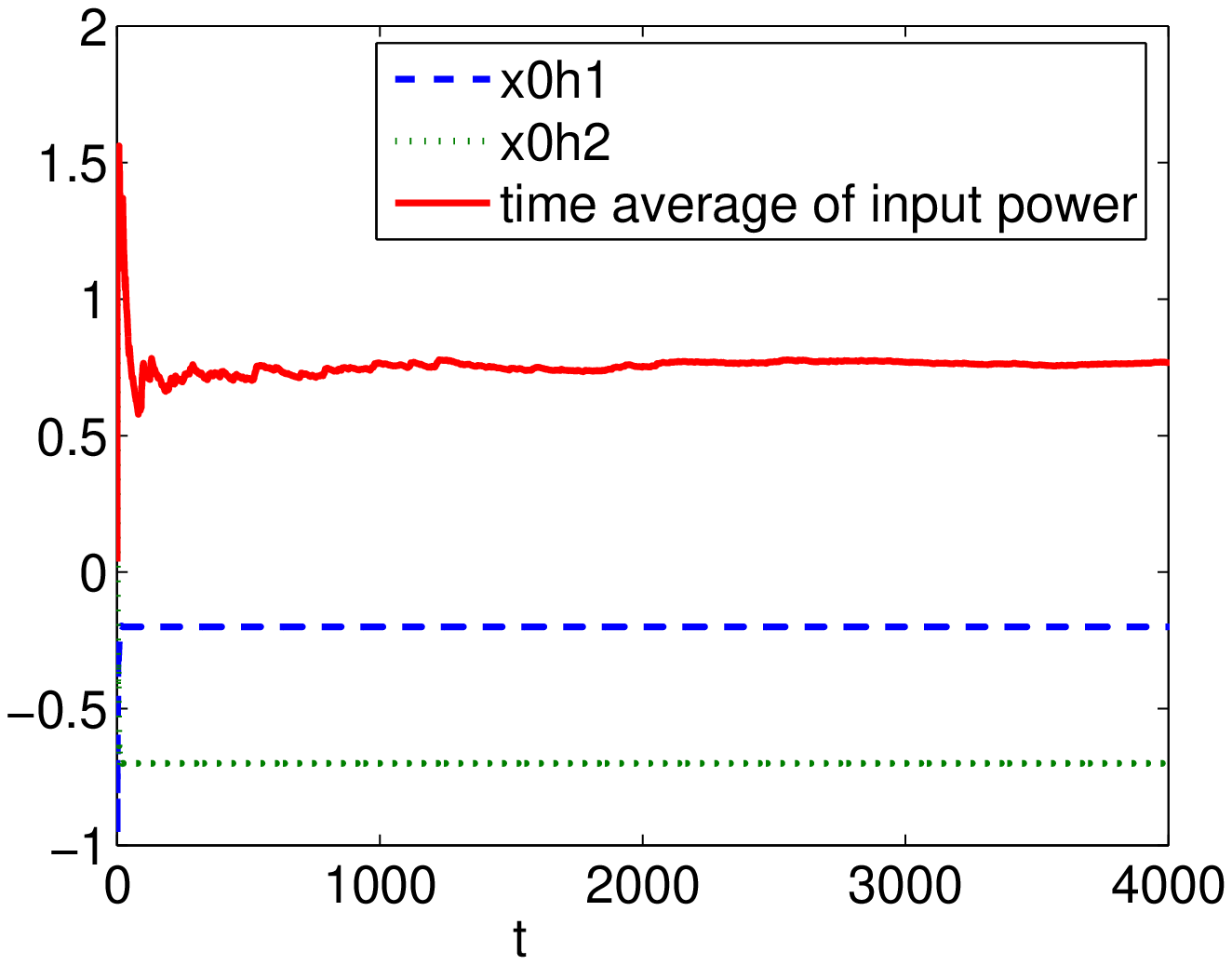}} }
\psfragscanon \psfrag{t}{\huge \textsf{number of channel uses}
\LARGE $(T+1)$} \subfigure[
]{{\scalebox{.45}{\includegraphics{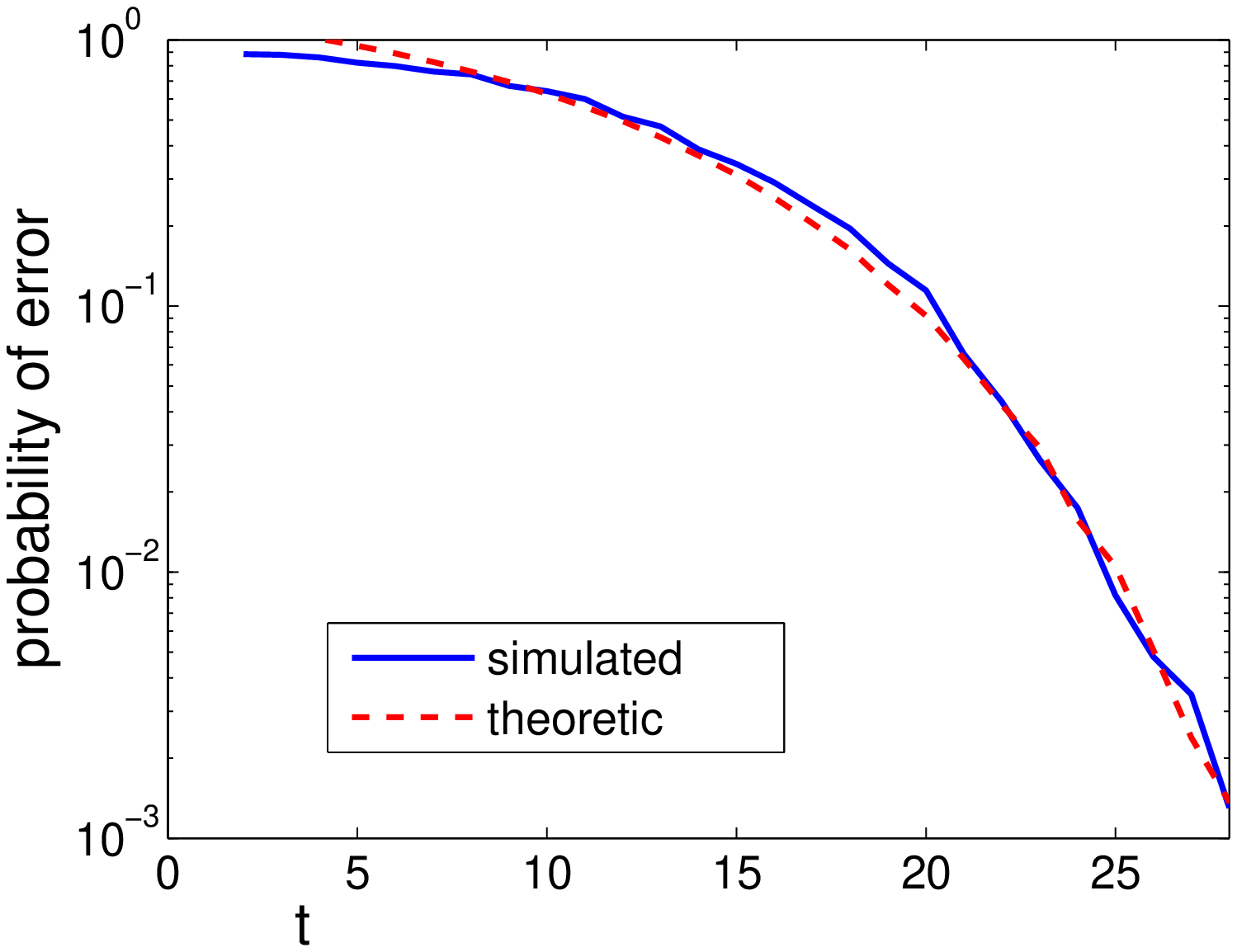}}}}

\caption{(a) Convergence of $\hatx_{0,t}$ to $x_0$, and
convergence of the average channel input power. (b) Simulated
probability of error and theoretic probability of error.
}\label{fig:simu}
\end{figure}

We report the simulation results using the modified communication
scheme with the optimal parameters given in (\ref{num:optal}).
Fig. \ref{fig:simu} (a) shows the convergence of $\hatx_{0,t}$ to
$x_0$, in which $x_0:=[-0.2,-0.7]'$.  Fig. \ref{fig:simu} (a) also
shows the time average of the channel input power, which converges
to the optimal power $P_\infty=0.743$. To compute the probability
of error, we let $\ep=0.2$, i.e., the signalling rate is equal to
$0.8 C_\infty$.  We demonstrate that this signalling rate is
achieved by showing that the simulated probability of error decays
to zero, see Fig. \ref{fig:simu} (b). Fig. \ref{fig:simu} (b) also
plots the theoretic probability of error computed from
(\ref{pet}), which is almost identical to the simulated curve.  In
addition, we see that the probability of error decays rather fast
within 28 channel uses.  The fast decay implies that the proposed
scheme allows shorter coding length and shorter coding delay; here
coding delay measures the time steps that one has to wait for the
message to be decoded at the receiver with small enough error
probability.

\section{Conclusions and future work}

We presented a coding scheme to achieve the asymptotic capacity
$C_\infty$ for a Gaussian channel with feedback. The scheme is
essentially the Kalman filter algorithm, and its construction
involves only a finite dimensional optimization problem.  We
established connections of feedback communication to estimation
and control. We have seen that concepts in estimation theory and
control theory, such as MMSE, CRB, minimum-energy control, etc.,
are useful in studying a feedback communication system. We also
verified the results by simulations.

Our ongoing research includes convexifying the optimization
problem (\ref{opt:dare}) to reduce the computation complexity, and
finding a more feasible scheme to fight against feedback noise
while keeping the feedback signal bounded. In future, we will
further explore the connections among information, estimation, and
control in more general setups (such as MIMO channels with
feedback).

\appendices

\section{Systems representations and equivalence} \label{appsec:equiv}

The concept of system representations and the equivalence between different
representations are extensively used in this paper.  In this subsection, we briefly
introduce system representations and the equivalence. For more thorough treatment, see
e.g. \cite{oppen:signals,chen:book,dahleh:book}.

\subsection{Systems representations} \label{appsub:represent}

Any discrete-time linear system can be represented as a linear mapping (or a linear
operator) from its input space to output space; for example, we can describe a
single-input single-output (SISO) linear system as
\be y^t = \calM_t u^t \ee
for any $t$, where $\calM_t \in \bbR^{(t+1) \times (t+1)}$ is the matrix
representation of the linear operator, $u^t \in \bbR^{t+1} $ is the stacked input
vector consisting of inputs from time 0 to time $t$, and $y^t \in \bbR^{t+1} $ is the
stacked output vector consisting of outputs from time 0 to time $t$.  For a (strictly)
causal SISO LTI system, $\calM_t$ is a (strictly) lower-triangular Toeplitz matrix
formed by the coefficients of the impulse response.  Such a system may also be
described as the (reduced) transfer function, whose inverse $z$-transform is the
impulse response; by a (reduced) transfer function we mean that its zeros are not at
the same location of any pole.

A causal SISO LTI system can be realized in state-space as
\be \left\{ \ba{lll} x_{t+1} &=& A x_t + B u_t \\
y_t &=& C x_t + D u_t, \ea \right. \label{state-space}\ee
where $x_t \in \bbR^l$ is the state, $u_t \in \bbR$ is the input, and $y_t \in \bbR$
is the output.  We call $l$ the \emph{dimension} or the \emph{order} of the
realization.  The state-space representation (\ref{state-space}) may be denoted as
$(A,B,C,D)$.   Note that in the study of input-output relations, it is sometimes
convenient to assume that the system is relaxed or at initial rest (i.e. zero input
leads to zero output), whereas in the study of state-space, we generally allow $x_0
\neq 0$, which is not at initial rest.  For multi-input multi-output (MIMO) systems,
linear time-varying systems, etc., see \cite{chen:book,dahleh:book}.

The state-space representation of an causal FDLTI system $\calM(z)$ is not unique.  We
call a realization $(A,B,C,D)$ \emph{minimal} if $(A,B)$ is controllable and $(A,C)$
is observable. All minimal realizations of $\calM(z)$ have the same dimension, which
is the minimum dimension of all possible realizations.  All other realizations are
called \emph{non-minimal}.

\textbf{An example}

\begin{figure}[h!]
\begin{center}
{\scalebox{.5}{\includegraphics{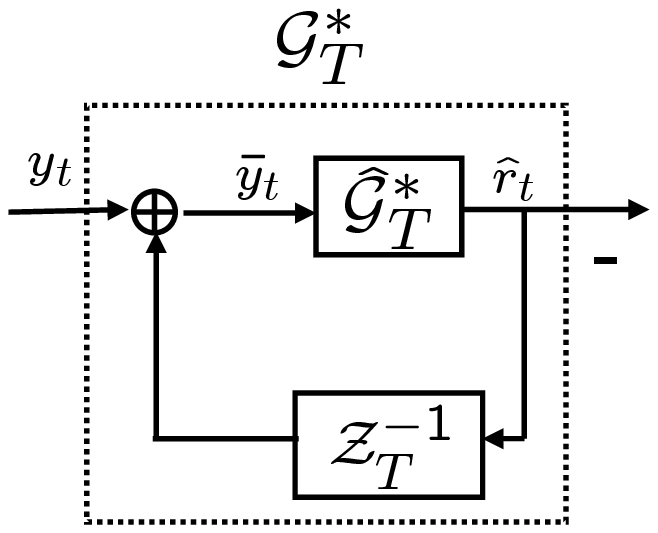}}} \caption{$\calG^*$ is a feedback connection of
$\hatG^*$ and $\calZ^{-1}$.} \label{fig:calG}
\end{center}
\end{figure}

We demonstrate here how we can derive a minimal realization of a system.  Consider
$\calG_T^*(A,C)$ in (\ref{eq:calGhatG}) in Section \ref{sec:kf}, which is given by
\be \calG_T^*(A,C) = - \hatG_T^* (I- \calZ_T^{-1} \hatG_T^*) ^{-1} , \label{eg:calG} \ee
where the state-space representations for $\hatG_T^*(A,C)$ and $\calZ_T^{-1}$ are
illustrated in Fig. \ref{fig:dual} (b) and Fig. \ref{fig:isicolor} (c).  Since
(\ref{eg:calG}) suggests a feedback connection of $\hatG^*$ and $\calZ^{-1}$ as shown in
Fig. \ref{fig:calG}, we can write the state-space for $\calG^*$ as
\be \left\{ \ba{lll}    \hatx_{t+1} &=& A \hatx_t + L_{1,t} e_t \\
                     \hatr_t &=& C \hatx_t\\
  \hat{\bar{s}}_{t+1} &=&  F \hat{\bar{s}}_{t} + G \hatr_t + L_{2,t} e_t\\
                        e_t &=& \bar{y}_t - H \hat{\bar{s}}_{t}  -\hatr_t \\
                        s_{a,t+1} &=& Fs_{a,t} + G \hatr_t \\
\bar{y}_t &=& y_t + Hs_{a,t} + \hatr_t. \ea \right. \ee
Then let $\hats_t:=\hat{\bar{s}}_{t} - s_{a,t}$, and we have
\be \left\{
\ba{lll}    \hatx_{t+1} &=& A \hatx_t + L_{1,t} e_t \\
                     \hatr_t &=& C \hatx_t\\
  \hats_{t+1} &=& F \hats_{t} + L_{2,t} e_t \\
                e_t &=& y_t - H \hats_t  . \ea \right. \ee
It is straightforward to check that this dynamics is controllable and observable, and
therefore it is a minimum realization of $\calG^*$.

\subsection{Equivalence between representations} \label{appsub:equiv}

\begin{definition} \label{def:equiv}

i) Two FDLTI systems represented in state-space are said to be \emph{equivalent} if
they admit a common transfer function (or a common transfer function matrix) and they
are both stabilizable and detectable.

ii) Fix $0 \leq T < \infty$.  Two linear mappings $\calM_{i,T}: \: \bbR^{ q(T+1) }
\rightarrow \bbR^{ p(T+1) }$, $i=1,2$, both at initial rest, are said to be
\emph{$T$-equivalent} if for any $u^T \in \bbR^{q(T+1)}$, it holds that
\be \calM_{1,T}(u^T) = \calM_{2,T}(u^T).\ee

\end{definition}

We note that i) is defined for FDLTI systems, whereas ii) is for general linear
systems. i) implies that, the realizations of a transfer function are not necessarily
equivalent. However, if we focus on all realizations that do not ``hide" any unstable
modes, namely all the unstable modes are either controllable from the input or
observable from the output, they are equivalent; the converse is also true.  ii)
concerns about the \emph{finite-horizon} input-output relations only. Since the states
are not specified in ii), it is not readily extended to infinite horizon: Any unstable
modes ``hidden" from the input and output will grow unboundedly regardless of input
and output, which is unwanted.

\textbf{Examples}

As we mentioned in Section \ref{subsec:isi}, for any $u^T$ and $N^T$, Fig.
\ref{fig:isicolor} (a) and (b) generate the same channel output $\tilde{y}^T$.  That
is, the mappings from $(u^T,N^T)$ to $\tilde{y}^T$ for the two channels are identical,
and both are given by
\be \tilde{y}^T = \calZ_T (\calZ_T^{-1} u^T + N^T ). \ee
Thus, we say the two channels are $T$-equivalent.

The feedback communication system (\ref{dyn:coding}), estimation system
(\ref{dyn:est}), and control system (\ref{dyn}) are $T$-equivalent, since for any
$N^T$, they generate the same innovations $e^T$.

\section{Finite-horizon: The feedback capacity and the CP structure} \label{app:cp}

\subsection{Feedback capacity $C_T$ } \label{appsub:c_t}

The following definition of feedback capacity is based on \cite{tati:capI}.

\begin{definition} \label{def:c_t}

The ``operational" or ``information" finite-horizon feedback capacity $C_T$, subject
to the average channel input power constraint
\be P_T:= \lim _{T \rightarrow \infty} \frac{1}{T+1} \E u^T{}' u^T \leq \mathcal{P},
\label{eq:pc}\ee
is
\be C_T(\calP):=C_T := \sup  \frac{1}{T+1}I(u^T \rightarrow y^T) , \label{opt:c_t}\ee
where $I(u^T \rightarrow y^T)$ is the directed information from $u^T$ to $y^T$, and
the supremum is over all possible feedback-dependent input distributions satisfying
(\ref{eq:pc}) and in the form
\be u_t=\gamma_t u^{t-1} + \eta_t y^{t-1} + \xi_t \label{u_app}\ee
for any $\gamma_t \in \bbR^{1 \times t}$, $\eta_t \in \bbR^{1 \times t}$, and
zero-mean Gaussian random variable $ \xi_t \in \bbR$ independent of $u^{t-1}$ and
$y^{t-1}$.

\end{definition}

\subsection{CP structure for colored Gaussian noise channel} \label{appsub:cpcolor}

We briefly review the CP coding structure for the colored Gaussian noise channel
specified in Section \ref{subsec:color}; see \cite{cover-pombra-1989,cover} for more
details of the CP structure. Let the colored Gaussian noise $Z^T$ have covariance
matrix $K_Z^{(T)}$, and
\be u^T := \calB_T Z^T +v^T, \label{eq:cpu}\ee
where $\calB_T$ is a $(T+1)\times(T+1)$ strictly lower triangular matrix, $v^T$ is
Gaussian with covariance $K_v^{(T)} \geq 0$ and is independent of $Z^T$.
\footnote{This $v^T$ is called innovations in \cite{cover,kavcic_it04}; it should not
be confused with the Kalman filter innovations in this paper.}  This generates channel
output
\be \tilde{y}^T = (I+\calB_T ) Z^T +v^T.  \label{eq:cpy}\ee
Then the highest rate that the CP structure can achieve in the sense of operational
and information is
\be \ba{lll} C_{T,CP}(\calP) &=& \disp \sup \frac{1}{T+1} I(v^T;\tilde{y}^T) \\
&=&  \disp \sup \frac{1}{2(T+1)} \log \frac{\det K_{\tilde{y}} ^{(T)} } {\det K_Z^{(T)}} \\
&=& \disp \sup \frac{1}{2(T+1)} \log \frac{ \det ((I+\calB_T) K_Z^{(T)}(I+\calB_T)' +
K_v^{(T)} ) } {\det K_Z^{(T)}}, \ea \label{cap:cpcolor} \ee
where the supremum is taken over all admissible $K_v^{(T)}$ and $\calB_T$ satisfying
the power constraint
\be P_T:=\frac{1}{T+1}\tr (\calB_T K_Z^{(T)}\calB_T'+K_v^{(T)}) \leq \calP.
\label{constr:cpcolor} \ee
Since the operational capacity definitions in \cite{cover-pombra-1989} and
\cite{tati:capI} coincide, we have $C_{T,CP}(\calP)=C_{T}(\calP)$.  This may also be
seen by observing that, any channel input (\ref{u_app}) can be rewritten in the form
of (\ref{eq:cpu}), but since (\ref{u_app}) is sufficient to achieve $C_T$, we conclude
that (\ref{eq:cpu}) is also sufficient to achieve $C_T$.

\subsection{CP structure for ISI Gaussian channel}

\begin{figure}[h!]
\begin{center}
{\scalebox{.5}{\includegraphics{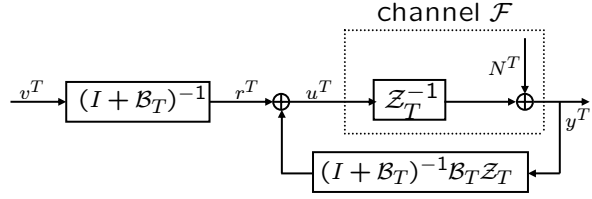}}} \caption{The block diagram of the CP
structure for ISI Gaussian channel $\calF$.} \label{fig:cpisi}
\end{center}
\end{figure}

By using the equivalence between the colored Gaussian noise channel and the ISI
channel $\calF$, we can derive the CP coding structure for $\calF$, which is obtained
from (\ref{eq:cpu}) by introducing a new quantity $r^T$ as
\be r^T : = (I+\calB_T)^{-1} v^T. \ee
By $Z^T=\calZ_T N^T$ and $\tilde{y}^T=\calZ_T y^T$, we have
\be \ba{lll} u^T &=& \calB_T \calZ_T N^T +(I+\calB_T) r^T \\
y^T &=& \calZ^{-1}_T (I+\calB_T) \calZ_T N^T + \calZ^{-1}_T (I+\calB_T) r^T \\
&=& \calZ^{-1}_T (I+\calB_T) (\calZ_T N^T + r^T) . \ea \label{eq:cpisi}\ee
This implies that, the channel input $u^T$ can be represented as
\be u^T =(I+\calB_T)^{-1} \calB_T \calZ_T y^T + r^T, \ee
which leads to the block diagram in Fig. \ref{fig:cpisi}.

The capacity $C_T$ now takes the form
\be \ba{lll} C_{T}(\calP) &= & \disp \sup \frac{1}{2(T+1)} \log \det K_y^{(T)}  \\
&=& \disp \sup \frac{1}{2(T+1)}  \log  \det \left(\calZ^{-1}_T (I+\calB_T) (\calZ_T
\calZ_T'
+ K_r^{(T)}) (I+\calB_T)' \calZ_T^{-1}{}' \right) \\
&=& \disp \sup \frac{1}{2(T+1)}  \log  \det  (\calZ_T \calZ_T' + K_r^{(T)})
\label{cap:cpisi} \ea \ee
where the supremum is over the power constraint
\be P_T:=\frac{1}{T+1}\tr (\calB_T \calZ_T \calZ_T' \calB_T'+(I+\calB_T) K_r^{(T)}
(I+\calB_T)' ) \leq \calP. \label{constr:cpisi} \ee
It is easily seen that the capacity in this form is identical to (\ref{cap:cpcolor}).

\subsection{Relation between the CP structure for ISI Gaussian channel and the general coding structure}
\label{appsub:cp_relation}

We can establish correspondence relationship between the CP structure for ISI Gaussian
channel $\calF$ in Fig. \ref{fig:cpisi} and the general coding structure for $\calF$
in Fig. \ref{fig:liu1}.  In fact, the general coding structure for $\calF$ in Fig.
\ref{fig:liu1} was initially motivated by the CP structure for channel $\calF$ in Fig.
\ref{fig:cpisi}.

For any fixed $(K_r^{(T)},\calB_T)$ in the CP structure, define in the general coding
structure that
\be \ba{lll} \calG_T &:=&(I+\calB_T)^{-1} \calB_T \calZ_T \\
A &:=& \Gamma_0^{-1} \left[ \begin{tabular}{c|c}0 & $I_T$ \cr \hline
* & * \end{tabular} \right] \Gamma_0  \;\; \in \bbR^{(T+1)\times(T+1)}\\
C &:=&  \left[ \matrix{1 &0& \cdots & 0} \right] \Gamma_0 ,\ea \ee
where $\Gamma_0 :=  (K_r^{(T)})^{\frac{1}{2}}$, and * can be any number.  (Note that
the case $K_r^{(T)} \geq 0$ but $K_r^{(T)}$ is not positive definite can be approached
by a sequence of positive definite $K_r^{(T)}$, and thus it is sufficient to consider
only positive definite $K_r^{(T)}$ in establishing the correspondence relation of the
two structures.) Then it is easily verified that $\calG_T$ is strictly lower
triangular, $(A,C)$ is observable with a nonsingular observability matrix
$\Gamma=\Gamma_0$, and $A$ can have eigenvalues not on the the unit circle and not at
the locations of $F$'s eigenvalues. Therefore, for any given $(K_r^{(T)},\calB_T)$, we
can find an admissible $(A,C,\calG_T)$, and it is straightforward to verify that they
generate identical channel inputs $u^T$.

Conversely, for any fixed admissible $(A,C,\calG_T)$ with $\in
\bbR^{(n+1)\times(n+1)}$, we can obtain an admissible $(K_r^{(T)},\calB_T)$ as
\be \ba{lll} \\
\calB_T &:=& \calG_T \calZ_T^{-1}(I-\calG_T \calZ_T^{-1}) ^{-1} \\
K_r^{(T)} &:=& \Gamma(A,C) \Gamma(A,C)', \ea \label{eq:g2b}\ee
which generates identical channel input $u^T$ as $(A,C,\calG_T)$ does.

As a result of the above reasoning, there is a corresponding relation between the CP
structure for $\calF$ and the general coding structure, and the maximum rate over all
admissible $(K_r^{(T)},\calB_T)$ (namely $C_T$) equals that over all admissible
$(A,C,\calG_T)$. In other words, we have

\begin{lemma} \label{lemma:ctn}
\be C_T(\calP)=C_{T,T}(\calP) .\ee
\end{lemma}

\proof Note that $C_{T,T}$ is the maximum rate over all admissible $(A,C,\calG_T)$
with $\in \bbR^{(T+1)\times(T+1)}$.  \endproof

This lemma implies that the general coding structure with an extra constraint $T=n$
becomes the CP structure, that is, in the CP structure, the dimension of $A$ is equal
to the horizon length.  One advantage of considering the general coding structure is
that we can allow $T \neq n$, which makes it possible to increase the horizon length
to infinity without increasing the dimension of $A$, a crucial step towards the Kalman
filtering characterization of the feedback communication problem.

Our study on the general coding structure also refines the CP structure.  We can now
identify more specific structure of the optimal $(K_v^{(T)},\calB_T)$.  Indeed, we
conclude that the CP structure needs to have a Kalman filter inside.  We may further
determine the optimal form of $\calB_T$. From (\ref{eq:g2b}) and (\ref{eq:calGhatG}),
we have that
\be \calB_T^* = - \hatG_T^*(A,C) \calZ_T^{-1} . \ee
Therefore, to achieve $C_T$ in the CP structure, it is sufficient to search
$(K_v^{(T)},\calB_T)$ in the form of
\be \ba{lll} K_v ^{(T)}&:=& (I- \hatG_T^*(A,C) \calZ_T^{-1} ) \Gamma(A,C) \Gamma(A,C)'  (I- \hatG_T^*(A,C) \calZ_T^{-1} ) '\\
\calB_T^* &:=& - \hatG_T^*(A,C) \calZ_T^{-1} . \ea \ee
Additionally, as $T$ tends to infinity, it can be easily shown
that $\{v_t\}$ is a stable process in order to achieve $C_\infty$.

\subsection{Proof of Proposition \ref{prop:kim}: Necessary condition for optimality}
 \label{appsub:kim}

In this subsection, we show that our general coding structure, in the form of
(\ref{dyn}), satisfies the necessary condition for optimality as presented in
Proposition \ref{prop:kim}.

Since $\{y_t\}$ is interchangeable with the innovations process $\{e_t\}$, in the
sense that they determine each other causally and linearly, it suffices to show that
$\E u_t e_\tau=0$. Note that
\be u_t = \bbD \bbX_t =\bbD \bbA \bbX_{t-1} - \bbD L_{t-1} e_{t-1}, \ee
and thus
\be  \ba{lll}\E u_t e_{t-1} &=& \E \bbD \bbA \bbX_{t-1} e_{t-1} - \bbD
L_{t-1} K_{e,t-1} \\
&\eqa& \E \bbD \bbA \bbX_{t-1}  \bbX_{t-1}' \bbC' + \E \bbD \bbA \bbX_{t-1}  N_{t-1} -
\bbD \bbA \Si_{t-1} \bbC' \\&=& \bbD \bbA \Si_{t-1} \bbC' + 0 - \bbD \bbA \Si_{t-1}
\bbC' =0, \ea \ee
where (a) follows from (\ref{dyn}) and (\ref{eq:Lt}). Similarly we can prove $\E u_t
e_\tau=0$ for any $\tau < t-1$.

\section{Infinite-horizon: The properties of the general coding structure} \label{app:inf}

\subsection{Proof of Proposition \ref{prop:dare}: Convergence to steady-state} \label{appsub:convg}

In this subsection, we show that system (\ref{dyn}) converges to a steady-state, as
given by (\ref{dyn:steady}).  To this aim, we first transform the Riccati recursion
into a new coordinate system, then show that it converges to a limit, and finally
prove that the limit is the unique stabilizing solution of the Riccati equation. The
convergence to the steady-state follows immediately from the convergence of the
Riccati recursion.

Consider a coordinate transformation given as
\be \ubbA:= \Phi \bbA \Phi^{-1}:=\left[\matrix{A & 0 \cr 0 & F}\right], \;\;\ubbC:=
\bbC \Phi^{-1}, \;\;\uSi_t:= \Phi \uSi_t \Phi',\ee
where
\be \Phi:=\left[\matrix{I_{n+1} & 0 \cr -\phi & I_m}\right],  \ee
and $\phi$ is the unique solution to the Sylvester equation
\be F \phi - \phi A = -GC. \ee
Note that the existence and uniqueness of $\phi$ is guaranteed by the assumption on
$A$ that $\lambda_i (-A) + \lambda_j (F) \neq 0$ for any $i$ and $j$ (see Section
\ref{sub:generalstructure}).

This transformation transforms $\bbA$ into block-diagonal form with the unstable and
stable eigenvalues in different blocks, and transforms the initial condition $\Si_0$
to
\be \uSi_0:=\Phi \left[\matrix{I_{n+1} & 0 \cr 0 & 0}\right] \Phi ' = \left[\matrix{I
& -\phi' \cr -\phi  & \phi \phi' }\right]. \ee
Therefore, the convergence of (\ref{ric:recur}) with initial condition $\Si_0$ is
equivalent to the convergence of
\be \uSi_{t+1} = \ubbA \uSi_t \ubbA' - \frac{\ubbA \uSi_t \ubbC' \ubbC \uSi_t \ubbA' }
{\ubbC \uSi_t \ubbC ' +1} \label{ric:recurNew} \ee
with initial condition $\uSi_0$.  By \cite{gallivan_riccati05}, $\uSi_t$ would
converge if
\be \det \left( \left[\matrix{0 & 0 \cr 0 & I_m}\right] - \uSi_0 \left[\matrix{I_{n+1}
& 0 \cr 0 & X_{22}} \right] \right) \not = 0, \ee
where $X_{22}$ is a positive semi-definite matrix (whose value does not affect our
result here).  Since
\be \ba{lll} \det \left( \left[\matrix{0 & 0 \cr 0 & I}\right] - \left[ \matrix{I &
-\phi' \cr -\phi  & \phi \phi' } \right] \left[\matrix{I & 0 \cr 0 & X_{22}}\right]
\right )
&=& \det \left( \left[\matrix{-I & \phi'X_{22}  \cr \phi & I-\phi \phi' X_{22}}\right]
\right)  \\
&=& \det (-I)  \det \left( I - \phi \phi'X_{22} + \phi \phi' X_{22} \right) \\
&\neq & 0, \ea \ee
we conclude that $\uSi_t$ converges to a limit $\uSi_\infty$.

This limit $\uSi_\infty$ is a positive semi-definite solution to
\be \uSi_\infty = \ubbA \uSi_\infty \ubbA' - \frac{\ubbA \uSi_\infty \ubbC' \ubbC
\uSi_\infty \ubbA' } {\ubbC \uSi_\infty \ubbC ' +1} . \label{ric:eqNew} \ee
By \cite{kailath:book}, (\ref{ric:eqNew}) has a unique stabilizing solution because
$(\ubbA,\ubbC)$ is observable and $\ubbA$ does not have any eigenvalues on the unit
circle.  Therefore, $\uSi_\infty$ is this unique stabilizing solution, which can be
computed from (\ref{ric:eqNew}) as (see also \cite{gallivan_riccati05})
\be \uSi_\infty = \left[ \matrix{\uSi_{11} & 0 \cr 0 & 0} \right] \ee
where $\uSi_{11}$ is the positive-definite solution to a reduced order Riccati
equation
\be \uSi_{11} = A \uSi_{11} A' - \frac{A \uSi_{11} (C+H\phi)' (C+H\phi) \uSi_{11} A' }
{(C+H\phi) \uSi_{11} (C+H\phi)' +1} . \label{ric:reduced} \ee
and has rank $(n+1)$ (cf. \cite{gallivan_riccati05}).  Thus, $\Si_t$ converges to
\be \Si_\infty = \left[ \matrix{\uSi_{11} & \uSi_{11}\phi' \cr \phi\uSi_{11} &
\phi\uSi_{11}\phi'} \right] \ee
with rank $(n+1)$.

\subsection{Infinite-horizon feedback capacities} \label{appsub:c_inf}

If the noise in the colored Gaussian channel forms a (an asymptotic) stationary
process, then $C_T(\calP)$ has a finite limit (cf. \cite{kim04}; the proof utilizes
the superadditivity of $C_T$, similar to the case of forward communication capacities
studied in \cite{gallager}), which also has the operational and information meanings.
Therefore, we have
\be \lim _{T \rightarrow \infty} C_T = C_\infty <\infty, \ee
where $C_\infty$ is the operational or information infinite-horizon capacity (cf.
\cite{cover-pombra-1989,tati:capI}).

By Lemma \ref{lemma:ctn}, the above implies that
\be \lim _{T \rightarrow \infty} C_{T,T} = C_\infty . \ee
Note that this does not simply lead to that $\lim _{n \rightarrow
\infty} \lim _{T \rightarrow \infty} C_{T,n} = C_\infty $ or
$C_\infty=C^s$, since we could not show that the involved limits
(including taking the supremum) are interchangeable in this case.

\section{Proof of Proposition \ref{prop:e0}: $K_\calE = 0$} \label{appsec:e0}

In this section, we prove that $K_\calE $ has to be 0 to ensure the optimality in
(\ref{opt:ykt}).

We first derive some properties of the communication system using the stationary GM
inputs and the steady-state Kalman filtering. The system dynamics is given by
\be \left\{ \ba{lll} u_t &=& d' \tildes_{s,t} + \calE_t\\
s_{t+1} &=& F s_t +G u_t\\
y_t &=& H s_t + N_t +u_t\\
\tilde{s}_{s,t+1}&=&s_t - \hats_{s,t} \\
\hats_{s,t+1} &=& F \hats_{s,t} + L_{s} e_t  \\
e_t &=& y_t - H \hats_{s,t} = (H+d') \tildes_{s,t} + \calE_t + N_t \\
\tildes_{s,t+1} &=& F \tildes_{s,t} + G u_t - L_{s} e_t ,  \ea \right.
\label{ykt:original}\ee
where $\hats_{s,0}=0$ and $\tildes_{s,0}=0$.  As before, the Kalman filter innovations
$\{e_t\}$ will play an important role.  The innovations process is white with variance
asymptotically equal to
\be K_e = 1+ K_{\calE} + (H+d') \Si_{s} (H+d')', \ee
where $\Si_{s}: = \E \tildes_s \tildes_s'$.  Following the same derivation for
Proposition \ref{prop:steady}, we know that the asymptotic information rate is given
by
\be  I(\calE;y) = \frac{1}{2} \log K_e, \ee
which is consistent with the result in \cite{kavcic_it04}.

\begin{figure}[h!]
\center \scalebox{.4}{\includegraphics{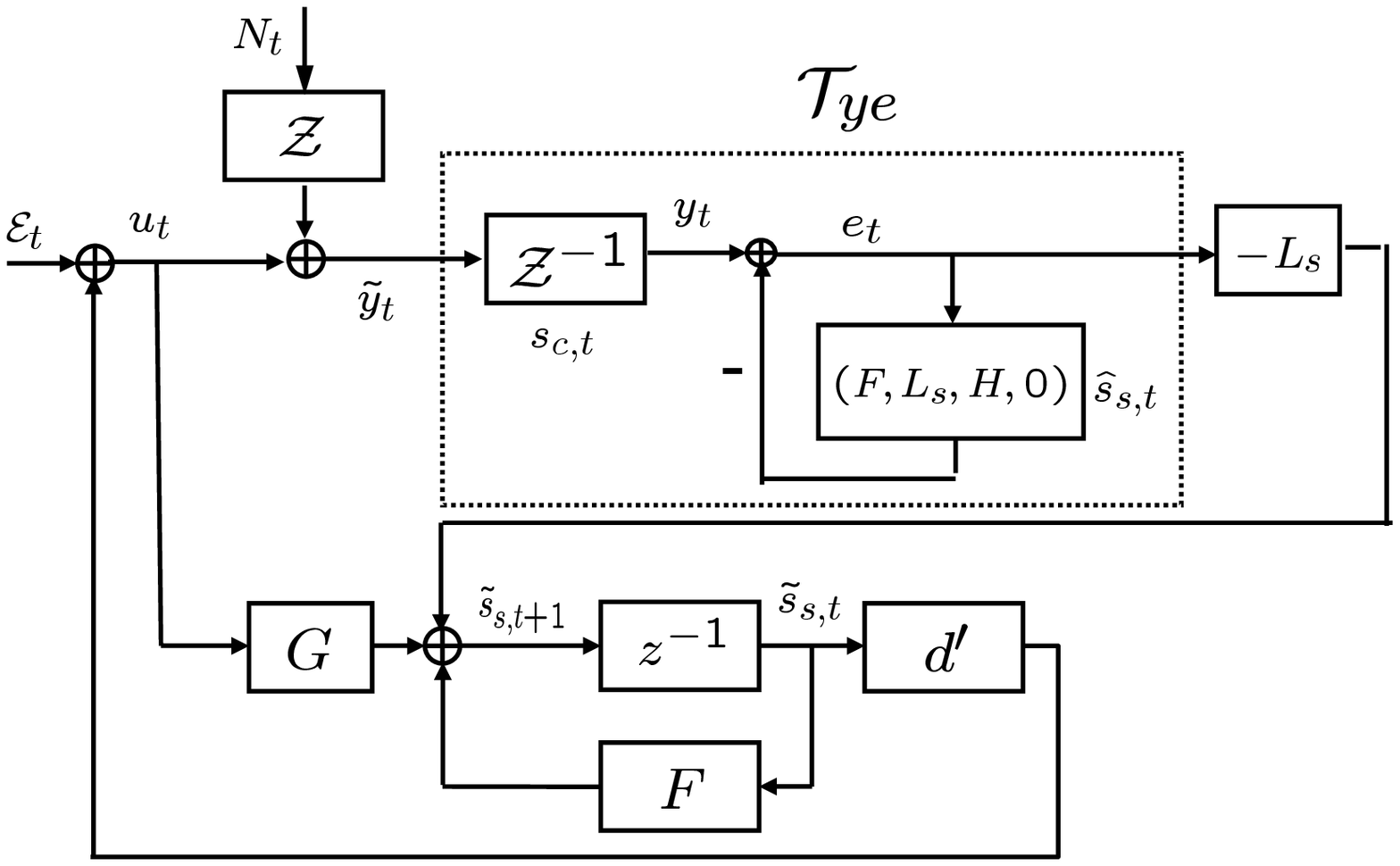}}  \caption{Block diagram for the
communication system using the GM inputs and Kalman filtering, where $s_{c,t}$ is the
state for $\calZ^{-1}$ with $s_{c,0}=0$, and $\hats_{s,t}$ is the state for system
$(F,L_s,H,0)$ with $\hats_{s,0}=0$.} \label{fig:ykt_ytilde}
\end{figure}

We now invoke the equivalence between the colored Gaussian channel and the ISI channel
$\calF$, that is, instead of generating $y$ by (\ref{ykt:original}), we generate $y$
by
\be \left\{ \ba{lll} \tildey_t &=& u_t+Z_t\\
s_{c,t+1} &=& F s_{c,t} + G \tildey_t\\
y_t &=& H s_{c,t} +\tildey_t , \ea \right. \ee
where $s_{c,0}=0$.  Since $Z^T = \calZ_T N^T$, the mapping from $(u,N)$ to $y$ here is
equivalent to that in (\ref{ykt:original}).  Therefore, (\ref{ykt:original}) becomes
\be \left\{ \ba{lll} u_t &=& d' \tildes_{s,t} + \calE_t\\
\tildey_t &=& u_t+Z_t\\
s_{c,t+1} &=& F s_{c,t} + G \tildey_t\\
y_t &=& H s_{c,t} +\tildey_t\\
\hats_{s,t+1} &=& F \hats_{s,t} + L_{s} e_t \\
e_t &=& y_t - H \hats_{s,t} = (H+d') \tildes_{s,t} + \calE_t + N_t \\
\tildes_{s,t+1} &=& F \tildes_{s,t} + G u_t - L_{s} e_t ,  \ea \right.
\label{ykt:original_ytilde}\ee
where $\hats_{s,0}=0$; see Fig. \ref{fig:ykt_ytilde} for the block diagram.

Our analysis of this system is facilitated by considering transfer functions.  Note
that
\be \ba{lll} \calT_{\calE u} &=& \bbS \\
\calT_{Nu} &=& \bbT \calZ , \ea \ee
where $\bbS$ is the sensitivity, and $\bbT:=\bbS-1$ is the complimentary sensitivity.
(The sensitivity $\bbS$ here should not be confused with the sensitivity in Section
\ref{subsec:dual}.) Then we have
\be \ba{lll}u &=& \bbS \calE + \bbT \calZ N \\
 \tilde{y}&=& \bbS (\calE + \calZ N)  . \ea \label{eq:yktff} \ee

Now assume that $d$ and $K_\calE$ form the \emph{optimal} solution to (\ref{opt:ykt}),
where $K_\calE \neq 0$, for contradiction purpose.  We can then compute the
corresponding optimal $\Si_s$, $L_s$, $\bbS$, $\bbT$, etc.  Fix the optimal $L_s$,
$\bbS$, and $\bbT$.  We will show that this leads to: 1) The whiteness of
$\{\tilde{y}_t\}$; 2) $L_{s}=G$; 3) $K_\calE = 0$ and hence contradiction.

\begin{figure}[h!]
\center {{\scalebox{.7}{\includegraphics{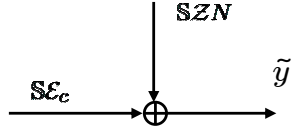}}}} \caption{ An equivalent forward
communication channel. Here $\bbS \calE_c$ is the effective input, $\bbS\calZ N$ is
the effective channel noise, and $\tilde{y}$ is the output.} \label{fig:ff}
\end{figure}

1) For fixed optimal values of $L_s$, $\bbS$, and $\bbT$, suppose that we can have the
freedom of choosing the power spectrum of $\calE$ in (\ref{ykt:original_ytilde}).
Since we have assumed the optimality of a white process $\{\calE_t\}$, it must hold
that any correlated process $\{\calE_{c,t}\}$ does not lead to a larger mutual
information than $\{\calE_t\}$ does.  Precisely, assume a stationary correlated
process $\{\calE_{c,t}\}$ replaces the white process $\{\calE_t\}$ in
(\ref{ykt:original_ytilde}).  Then $\{\calE_t\}$ yields the maximum achievable rate
over all possible $\{\calE_{c,t}\}$, i.e., it solves
\be \label{opt:I4calE}
\begin{array}{cl}
& \displaystyle{ \max_{ L_s,\bbS,\bbT \textrm{ \small fixed}, \calS_{\calE_c} (e^{j
2\pi \theta})}}
 \displaystyle I(\calE_c; \tildey) .\\
      &^{ s.t. \: \E u^2  \leq \calP}
\end{array} \ee
Since
\be I(\calE_c; \tildey)=h(\tildey) - h(\tildey|\calE_c) = h(\tildey) -h(\bbS\calZ N)
\ee
and $h(\bbS\calZ N)$ is fixed for fixed $\bbS$, the above optimization is equivalent
to
\be \label{opt:ff1}
\begin{array}{cl}
& \displaystyle{ \max_{ \calS_{\calE_c} (e^{j 2\pi \theta})}}
 \displaystyle \frac{1}{2} \displaystyle \int_{-\frac{1}{2}}
^{\frac{1}{2}} \log \calS_{\tilde{y}} (e^{j 2\pi \theta}) d
\theta .\\
      &^{ s.t. \:  \E u^2 = \int_{-\frac{1}{2}}
^{\frac{1}{2}} \calS_{\bbS} (e^{j 2\pi \theta}) \calS_{\calE_c} (e^{j 2\pi \theta}) +
\calS_{\bbT} (e^{j 2\pi \theta}) \calS_{\calZ} (e^{j 2\pi \theta}) d \theta \leq
\calP}
\end{array} \ee
However, this optimization problem is equivalent to solving, for some $\calP_1 \geq
0$,
\be \label{opt:ff2}
\begin{array}{cl}
& \displaystyle{ \max_{ \calS_{\calE_c} (e^{j 2\pi \theta})}}
 \displaystyle \frac{1}{2} \displaystyle \int_{-\frac{1}{2}}
^{\frac{1}{2}} \log \left( \calS_{\bbS} (e^{j 2\pi \theta}) \calS_{\calE_c} (e^{j 2\pi
\theta}) + \calS_{\bbS} (e^{j 2\pi \theta}) \calS_{\calZ} (e^{j 2\pi \theta}) \right)
d
\theta ,\\
      &^{ s.t. \:  \int_{-\frac{1}{2}}
^{\frac{1}{2}} \calS_{\bbS} (e^{j 2\pi \theta}) \calS_{\calE_c} (e^{j 2\pi \theta})  d
\theta \leq \calP_1 }
\end{array} \ee
which we identify as a new forward communication problem, see Fig. \ref{fig:ff}. In
this problem, we want to tune the power spectrum of $\bbS \calE_c$, the effective
channel input, to get the maximum rate. The optimal solution is given by waterfilling,
namely, the power spectrum $\calS_{\bbS} (e^{j 2\pi \theta}) \calS_{\calE_c} (e^{j
2\pi \theta})$ needs to waterfill the power spectrum $\calS_{\bbS} (e^{j 2\pi \theta})
\calS_{\calZ} (e^{j 2\pi \theta})$.  By optimality of $\{\calE_t\}$, $K_\calE
\calS_{\bbS} (e^{j 2\pi \theta}) $ is the waterfilling solution.

Since $ \calS_{\bbS} (e^{j 2\pi \theta}) =0$ for some $\theta$ if and only if
$\bbS(z)$ has a zero for that $\theta$ on the unit circle, and since $\bbS(z)$ is a
finite dimension transfer function with a finite number of zeros, the power spectrum
$\calS_\bbS (e^{j 2\pi \theta})$ cannot have zero amplitude at any interval.  This
follows that the support of the channel input spectrum $K_\calE \calS_{\bbS} (e^{j
2\pi \theta}) $ is $[-1/2,1/2]$.

In waterfilling, if the support of input spectrum is $[-1/2,1/2]$, then the output
spectrum must be flat.  This is easily proven by contradiction.  Thus,
$\{\tilde{y}_t\}$ is a white process.  Let us assume that its variance is $\s^2$.

\begin{figure}[h!]
\center {{\scalebox{.5}{\includegraphics{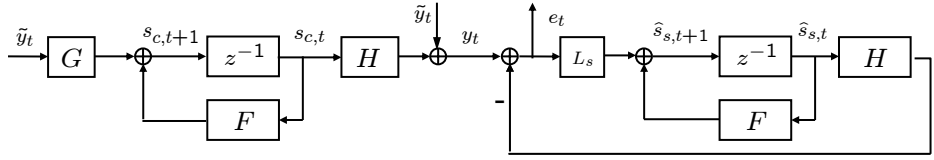}}}} \caption{ The state-space
representation of the transfer function $\calT_{ye}$.} \label{fig:tye}
\end{figure}

2) Note that both $\tilde{y}$ and $e$ have white spectrum, which imposes condition on
the choice of $L_s$.  The transfer function $\calT_{ye}$ is illustrated in Fig.
\ref{fig:tye}, where we can see that its structure is a Kalman filter structure.  To
make $e$ white, it is necessary to choose $L_s$ to be the Kalman filter gain (cf.
\cite{kailath:book}), given by
\be L_s := \frac{F \Si_c H' + \sigma^2 G}{H \Si_c H' + \sigma^2}, \ee
where $\Si_c$ is the estimation error covariance matrix and is a nonnegative solution
to the Riccati equation
\be \Si_c = F \Si_c F' +\sigma^2 GG' - \frac{(F \Si_c H' + \sigma^2 G) (F \Si_c H' +
\sigma^2 G)' }{H \Si_c H' + \sigma^2} .\ee
Clearly, $\Si_c=0$ is a solution to the Riccati equation.  By \cite{kailath:book}, it
is also the unique nonnegative solution. Hence, we need to choose $L_s:=G$.

3)  The fact that $L_s=G$ leads to reduction of system (\ref{ykt:original_ytilde}) or
equivalently (\ref{ykt:original}). We have
\be \ba{lll} \tilde{s}_{t+1} &=& (F-GH) \tilde{s}_t -G N_t \\
\Si_s &=& (F-GH)\Si_s(F-GH)' +GG' . \ea \ee
In the case that $(F-GH)$ is unstable, the closed-loop of (\ref{ykt:original_ytilde})
is unstable and cannot transmit information.  In the case that $(F-GH)$ is stable, the
steady-state of $\Si_s$ depends only on $(F,G,H)$ and is \emph{independent} of the
choice of $d$ and $K_\calE$, and thus (\ref{opt:ykt}) becomes
\be \label{opt:ykt1}
\begin{array}{cl}
C_\infty =& \displaystyle{  \max_{\Si_s \textrm{ \small fixed},  d \in \bbR^{m},
K_\calE \in \bbR}}
 \displaystyle \frac{1}{2} \log (1+ K_\calE + (H+d') \Si_s
(H+d')'). \\
      &^ {s.t.\: \calP = d' \Si_s d + K_\calE}
\end{array} \ee
This is equivalent to
\be \label{opt:ykt2}
\begin{array}{cl}
&\displaystyle{ \max_{d \in \bbR^{m}, K_\calE \in \bbR}}
\displaystyle H\Si_sd, \\
      &^{s.t. \: d' \Si_s d \leq \calP - K_\calE}
\end{array} \ee
which requires $K_\calE=0$.

\section{Optimality of the proposed coding scheme} \label{appsec:achievImax}

\subsection{Proof of Proposition \ref{prop:fdim}: Finite dimensionality of the optimal
scheme} \label{appsub:fdim}

i) To show that $C_{\infty,n}$ is non-decreasing as $n$ increases, note that, an encoder
$(A,C)$ of dimension $(n+1)$ can be arbitrarily approximated by a sequence of encoders $
\{(A_i,C_i)\}$ of dimension $(n+2)$ in the form of
\be \left( \left[ \ba{c|c}A  & 0 \cr \hline 0 & 1 \ea \right],\left[ \ba{c|c} C &
 \frac{1}{i} \ea \right]\right), \ee
and therefore the supremum in (\ref{def:c_infn}) with encoder dimension $(n+2)$ is no
smaller than the supreme with encoder dimension $(n+1)$.  So $C_{\infty,n}$ is increasing
in $n$.

ii) By proposition \ref{prop:steady} and the definition for $C_{\infty,m-1}(\calP)$, the
optimization problem for solving $C_{\infty,m-1}(\calP)$ is given by
\begin{equation}
\begin{array}{lcl}
  C_{\infty,m-1}(\calP) = &
   \disp \sup_{A \in \bbR^{m \times m},C}  & \disp \frac{1}{2}\log (\bbC \Si \bbC' +1)  \\
     &  ^{s.t. \Sigma = \bbA \Sigma \bbA' - \bbA \Sigma \bbC' (\bbC \Sigma
\bbC ' +1)^{-1} \bbC \Sigma \bbA' } \\
& \;\; ^{ \bbD \Sigma \bbD' =\calP}
\end{array} \label{opt:dare_fdim1}
\end{equation}
To compare it with $C_\infty(\calP)$, we rewrite (\ref{opt:ykt})
and (\ref{opt:ykt_constr}) in another form, incorporating
$K_\calE=0$.  Define
\be \ba{lll} \bar{\bbA}&:=& \left[ \ba{c | c} F+Gd' & 0 \\ \hline Gd' & F \ea \right] \\
\bar{\bbC}&:=& [d \quad H] \\
\bar{\bbD}&:=&  [d \quad 0]  \\
\bar{\Si} &:=& \left[ \ba{c | c} \Si_s & \Si_s \\ \hline \Si_s & \Si_s \ea \right]. \ea
\ee
It is then straightforward to verify that
\be \ba{rcl} \disp \frac{1}{2} \log (1+ (H+d') \Si_s (H+d')') &=& \disp \frac{1}{2}\log
(1+\bar{\bbC} \bar{\Si} \bar{\bbC}' ) \\
d' \Si_s d &=& \bar{\bbD} \bar{\Sigma} \bar{\bbD}' \\
\disp \bar{\bbA} \bar{\Sigma} \bar{\bbA}' - \bar{\bbA} \bar{\Sigma} \bar{\bbC}'
(\bar{\bbC} \bar{\Sigma} \bar{\bbC} ' +1)^{-1} \bar{\bbC} \bar{\Sigma} \bar{\bbA}'  &=&
\bar{\Sigma}, \ea \ee
which yields that
\begin{equation}
\begin{array}{lcl}
  C_\infty(\calP) = & \displaystyle{
   \sup_{d \in \bbR^{m }}}  & \disp \frac{1}{2}\log (1+\bar{\bbC} \bar{\Si} \bar{\bbC}' )  \\
     &  _{s.t. \bar{\Sigma} =\bar{\bbA} \bar{\Sigma} \bar{\bbA}' - \bar{\bbA} \bar{\Sigma} \bar{\bbC}' (\bar{\bbC}
\bar{\Sigma} \bar{\bbC} ' +1)^{-1} \bar{\bbC} \bar{\Sigma} \bar{\bbA}' }   \\
& \;\; _{ \bar{\bbD} \bar{\Sigma} \bar{\bbD}' =\calP}
\end{array} \label{opt:dare_fdim2}
\end{equation}
Comparing (\ref{opt:dare_fdim2}) with (\ref{opt:dare_fdim1}), we
conclude that $C_{\infty,m-1}(\calP) \geq C_\infty(\calP)$.
However, since for each $(A,C)$, the channel input sequence is
stationary by the steady-state characterization of the general
coding structure, it holds that $C_{\infty,m-1}(\calP) \leq
C_\infty(\calP) $. Therefore, we have
\be C_{\infty,m-1}(\calP) = C_\infty(\calP) .\ee
Then ii) follows from i) immediately.

\subsection{Proof of Proposition \ref{prop:IeqImax}:
Achieving $C_\infty$ in the information sense}
\label{appsub:IeqImax}

By Proposition \ref{prop:fdim}, the optimization problem for
solving $P_\infty (\calR)$ in (\ref{opt:imax_pmin}) (which is
equivalent to solving $C_\infty(\calP)$) can be reformulated as
\begin{equation}
\begin{array}{lcl}
[A^{opt},C^{opt},\Si^{opt}]:=
   & \displaystyle{\textnormal{arg} \inf_{A \in \bbR^{m \times m},C}}  &  \bbD \Sigma \bbD',  \\
     &  ^{s.t. \Sigma = \bbA \Sigma \bbA' - \bbA \Sigma \bbC' (\bbC \Sigma
\bbC ' +1)^{-1} \bbC \Sigma \bbA' } \\
& \;\; ^{\log DI(A)=\calR}
\end{array} \label{opt:dare1}
\end{equation}
for any desired rate $\calR$. Without loss of generality, we may assume that $(A,C)$
is in the observable canonical form, i.e.
\be \ba{lll} A :=\left[ \ba{c|c} 0_{n \times 1} & I_n \cr \hline
 a_n & a_{n-1} \cdots a_1
\ea \right]  \\
C :=\left[ \ba{cc} 1& 0_{1 \times n} \ea \right] . \ea \label{AC:ctrb}\ee
Observe that $\det A=a_n$.  Thus, $DI(A)=|\det A|=|a_n|$ if $A$ does not contain
stable eigenvalues, and $DI(A)>|\det A|=|a_n|$ otherwise.

As a consequence, if we search over $A$ with $a_n$ fixed to be
$2^\calR$ or $-2^\calR$, we actually enforce $DI(A) \geq 2^R$.
However, the optimal solution must satisfy $DI(A^{opt})=2^\calR$,
since otherwise the system has a rate equal to
$R_{\infty,m-1}=\log DI(A^{opt}) > \calR$, which would require
more power than the case that $R_{\infty,m-1} = \calR$; notice
that (\ref{opt:dare1}) is a power minimization problem.  To
summarize, we can remove the constraint $\log DI(A)=\calR$ by
letting $a_n = \pm 2^\calR$ in (\ref{AC:ctrb}), and the optimal
solution $A$ does not contain stable eigenvalues.  Furthermore,
note that unit-circle eigenvalues do not generate any rate or
power and hence can be removed.  Thus, if $A^{opt}$ has $(n^*+1)$
unstable eigenvalues, we can solve the optimization problem with
$A$ having size $(n^*+1)$ and the obtained optimal solution still
achieves $C_\infty$.

\subsection{Proof of Proposition \ref{prop:main_RD}: Optimality in the analog transmission}
 \label{appsub:main_RD}

The end-to-end distortion is given by
\be \ba{lll}\MSE(\hatW_t) &=& \E (W-\hatW_t)(W-\hatW_t)' \\
& =& \E (x_0-\hatx_{0,t})(x_0-\hatx_{0,t})' \\
& =& \E (A^{-t-1} x_{t+1}-A^{-t-1}\hatx_{t+1}) (A^{-t-1} x_{t+1}-A^{-t-1}\hatx_{t+1})' \\
& =& \E A^{-t-1} \tildex_{t+1} \tildex_{t+1}' A'{}^{-t-1} \\
& =&  A^{-t-1} \Si_{x,t+1}  A'{}^{-t-1} , \ea \label{MSE_RD} \ee
where
\be \Si_{x,t+1} : = [I,0] \Si_{t+1} [I,0]' \ee
and the expectation is w.r.t. the randomness in $W$ and $\hatW_t$.  By rate-distortion
theory, the above distortion needs an asymptotic rate $R$ satisfying
\be \ba{lll} R &\geq & \disp \lim _{t \rightarrow \infty} \frac{1}{2(t+1)} \log \frac{\calP^{n^*+1}}{\det MSE(\hatW_t)} \\
&= & \disp \lim _{t \rightarrow \infty} \frac{1}{2(t+1)} \log \frac{\det A ^{2t+2}} {\det \Si_{x,t+1}} \\
&= & \disp  \log |\det A| .  \ea \label{RDineq}\ee
From Proposition \ref{prop:IeqImax}, we know that $\log |\det
A^*|$ equals $C_\infty$ and the average channel input power equals
$\calP$.  Because $C_\infty$ is the supremum of asymptotic rate,
it follows that the equality in (\ref{RDineq}) is achieved.  Then
we see that the proposition holds.

\subsection{Proof of Proposition \ref{prop:main}: Optimality in digital transmission}
\label{appsub:main}

It is sufficient to show that $R_{\infty,n} (A,C)$ is achievable for any fixed
$(A,C)$.  To show this, for the fixed $(A,C)$, construct the scheme in Fig.
\ref{fig:liu1} and use $\calG_T^*$, the Kalman-filter based optimal receiver.  The
closed-loop (\ref{dyn}) is stabilized  and will converge to its steady-state for large
enough $T$.

We can then directly verify that Theorems 4.3 and 4.6 in \cite{elia_c5} are applicable
to the (steady-state) LTI system. These theorems assert that, if the closed-loop
system is stabilized, then we can construct a sequence of codes to reliably (in the
sense of vanishing probability of error) transmit the initial conditions associated
with the open-loop unstable eigenvalues of $A$ (denoted $a_0,\cdots,a_k$, if any), at
a rate
\be R := (1-\epsilon) R_{\infty,n}(A,C)  \ee
for any $\epsilon >0$, and in the meantime, $P_{\infty,n}(A,C) \leq \mathcal{P}$
holds. Therefore, we conclude that, for any $(A,C)$, the portion of $W$ that is
associated with the unstable eigenvalues of $A$ is transmitted reliably from the
transmitter to the receiver at rate arbitrarily close to $R_{\infty,n}(A,C)$.
Moreover, we notice that we can achieve $C_{\infty,n}$ by a sequence of purely
unstable $(A,C)$ (i.e. $k=n$), in which the initial condition $W$ is the message being
transmitted. This follows that $W$ is transmitted at the capacity rate.

In addition, \cite{elia_c5} showed that for any choice of $x_0$, it holds that
\be PE_T= 1- \prod_{i=0}^n \left(1- 2Q \left(\frac{\sigma_{T,i}^{-\ep }}{2} \right)
\right), \label{pet} \ee
where $\sigma_{T,i}$ is the square root of the $i$th eigenvalue of $\MSE(\hatx_{0,T})$,
and
\be \ba{lll}\MSE(\hatx_{0,T}) &=& \E (x_0-\hatx_{0,T}) (x_0-\hatx_{0,T})' \\
& =&  A^{-T-1} \Si_{x,T+1}  A'{}^{-T-1} . \ea \ee
Note that the expectation is w.r.t. the randomness in $\hatx_{0,T}$ only, different
from (\ref{MSE_RD}), and that asymptotically $\Si_{t+1}$ and hence $\Si_{x,T+1}$ are
independent on the choice of $x_0$.

It then holds for each $i$,
\be \ba{lll} (\sigma_{T,i})^2 &\leq& \lambda_{\max} (\MSE(\hatx_{0,T})) \\
&=& \lambda _{\max} (A^{-T-1} \Si_{x,T+1}  A'{}^{-T-1}) \\
&\eqa& \lambda _{\max} ( A'{}^{-T-1}  A{}^{-T-1} \Si_{x,T+1} ) \\
&\leqb & \bar{\sigma} ( A'{}^{-T-1}  A{}^{-T-1} \Si_{x,T+1} ) \\
&\leqc & \bar{\sigma} ( A'{}^{-T-1}  A{}^{-T-1} ) \bar{\sigma} ( \Si_{x,T+1} ) \\
&= & \left(\bar{\sigma} ( A' A )\right)^{-T-1} \bar{\sigma} (\Si_{x,T+1})
 \ea \label{ineq:sigma} \ee
where $\lambda_{\max} (M)$ denotes the maximum eigenvalue of $M$, $\bar{\sigma} (M)$
denotes the maximum singular value of $M$, (a) follows from $\lambda(AB)=\lambda(BA)$,
(b) follows from $|\lambda(A)| \leq \bar{\sigma}(A)$, and (c) is because the maximum
singular value is an induced norm.  Since $\Si_{x,T+1}$ converges to steady-state
value exponentially, the above implies that, for $T$ large enough, each $\sigma_{T,i}$
decays to zero exponentially as $T$ increases.

Now using the union bound and the Chernoff bound, we have
\be \ba{lll} PE_T & \leq& \disp \sum_{i=0}^n 2 Q \left(\frac{\sigma_{T,i}^{-\ep
}}{2} \right) \\
&\leq&  \disp \sum_{i=0}^n \frac{4}{\sqrt{2 \pi} \sigma_{T,i}^{-\ep }  } \exp \left(-
\frac{\sigma_{T,i}^{-2\ep }}{8}\right) , \ea \ee
and hence $PE_T$ decreases to zero doubly exponentially since $\ep>0$ and
$\sigma_{T,i}$ decays exponentially.  Thus we prove the proposition.

\vspace{.25in} \hspace*{2.5in} \textbf{ACKNOWLEDGEMENTS} \vspace{.15in}

The authors would like to thank Anant Sahai, Sekhar Tatikonda, Sanjoy Mitter, Zhengdao
Wang, Murti Salapaka, Shaohua Yang, Donatello Materassi, and Young-Han Kim for useful
discussion.

\vspace{-15pt} \hfill \markright{\textsf{References}} \small
\bibliographystyle{unsrt}

\end{document}